\documentclass[12pt,preprint]{aastex}
\slugcomment{To be submitted to The Astrophysical Journal}
\shorttitle{On the Spiral Structure of \objectname[]{NGC~2915} and Dark Matter}
\shortauthors{Masset \& Bureau}
\received{2002 July 4}
\begin{document}
\newcommand{\msun}{$M_{\mbox{\scriptsize $\sun$}}$}
\newcommand{\coltwo}[2]{\mbox{$\displaystyle\mbox{#1}\atop{\displaystyle\mbox{#2}}$}}
\title{On the Spiral Structure of \objectname[]{NGC~2915} and Dark Matter}
\author{F.\ S.\ Masset\altaffilmark{1}}
\affil{Service d'Astrophysique, CE-Saclay, 91191 Gif/Yvette Cedex,
 France}
\email{fmasset@cea.fr}
\and
\author{M.\ Bureau\altaffilmark{2}}
\affil{Columbia Astrophysics Laboratory, 550~West 120th~Street, 1027
 Pupin Hall,\\MC~5247, New York, NY~10027}
\email{bureau@astro.columbia.edu}
\altaffiltext{1}{Send offprint requests to F.\ S.\ Masset: fmasset@cea.fr}
\altaffiltext{2}{Hubble Fellow}
\begin{abstract}
\objectname[]{NGC~2915} is a blue compact dwarf galaxy embedded in an
extended, low surface brightness \ion{H}{1} disk exhibiting a
two-armed spiral structure and a central bar-like component. Commonly
accepted mechanisms are unable to explain the existence of these
patterns and Bureau et al.\ proposed disk dark matter (scaling with
the \ion{H}{1} distribution) or a rotating triaxial dark halo as
alternative solutions. In an attempt to explore these mechanisms,
hydrodynamical simulations were run for each case and compared to
observations using customized column density and kinematic
constraints. The spiral structure can be accounted for both by an
unseen bar or triaxial halo, the former fitting the observations
slightly better. However, the large bar mass or halo pattern frequency
required make it unlikely that the spiral wave is driven by an
external perturber. In particular, the spin parameter $\lambda$ is
much higher than predicted by current cold dark matter (CDM) structure
formation scenarios. The massive disk models show that when the
observed gas surface density is scaled up by a factor about $10$, the
disk develops a spiral structure resembling closely the observed one,
in perturbed density as well as perturbed velocity. This is consistent
with more limited studies in other galaxies and suggests that the disk
of \objectname[]{NGC~2915} contains much more mass than is visible,
tightly linked to the neutral hydrogen. A classic (quasi-)spherical
halo is nevertheless still required, as increasing the disk mass
further to fit the circular velocity curve would make the disk
violently unstable. Scaling the observed surface density profile by an
order of magnitude brings the disk and halo masses to comparable
values within the disk radius. The surface density remains under
Kennicutt's star formation threshold for a gaseous disk and no stars
are expected to form, as required by observations.
\end{abstract}
\section{Introduction\label{sec:intro}}

\objectname[]{NGC~2915} is a blue compact dwarf (BCD) galaxy at a
distance $D=5.3\pm1.6$~Mpc \citep*{mmc94}. In the optical, it displays
a high surface brightness blue core and a red diffuse population. The
core is the locus of high mass star formation and possesses a high
excitation, low metallicity \ion{H}{2} region spectrum; the red
population has an exponential profile with a low extrapolated central
surface brightness \citep{mmc94}. The radio properties of
\objectname[]{NGC~2915} are rather extreme, with an \ion{H}{1} disk
extending to 22 radial scalelengths in the $B$-band \citep[5 Holmberg
radii;][]{mcbf96}. This disk displays a short central bar overlapping
the optical emission and a well-developed outer two-armed spiral
extending to its edge (see Fig.~\ref{fig:iv_images}). 
The signature of an oval distortion is also
present in the velocity field. Standard modeling of the derived
circular velocity curve yields a blue mass-to-light ratio ${\cal
M}_T/L_B\ga65$ at the last measured point (${\cal M}_{\rm dark}/{\cal
M}_{\rm luminous}\approx19$), making \objectname[]{NGC~2915} one of
the darkest disk galaxies known \citep{mcbf96}. The core of the dark
matter halo also appears unusually dense and compact. In fact, the
stellar content can be neglected at all radii without greatly
affecting the fit. The optical properties of \objectname[]{NGC~2915},
discussed in more detail by \citet{mmc94}, are thus unimportant for
this work. We will rather focus on its unique \ion{H}{1} properties,
described more fully in \citet{mcbf96}.

\citet{bfpm99} discussed the origin of the \ion{H}{1} bar and spiral
pattern in \objectname[]{NGC~2915}. They argue for a common, slow
pattern speed $\Omega_p=8.0\pm2.4$~km~s$^{-1}$~kpc$^{-1}$, yielding a
corotation radius to bar semi-length ratio $r_{\rm
cr}/r_b\ga1.7$ ($r_b=180\arcsec=4.6$~kpc). This lower limit is reached
for $\Omega_p=10.4$~km~s$^{-1}$~kpc$^{-1}$, while the limited extent of
the circular velocity curve prevents the derivation of a proper upper
limit. Although \citeauthor{bfpm99}'s (\citeyear{bfpm99}) measurement
using \citet{tw84a} method 
is plagued by uncertainties, it is consistent with the idea that bars
in dense halos should be slow \citep[e.g.][]{ds98}. Furthermore, the
low (luminous) disk surface density and the locations of the
pseudo-rings make it unlikely that swing amplification \citep{t81} or
bar-driving \citep{s81,s84,brsbc94} can explain the bar and spiral
patterns. Because \objectname[]{NGC~2915} is isolated, gravitational
interactions cannot be invoked as an exciting mechanism either
\citep{n87}. \citet{bfpm99} proposed two alternatives: i) the
\ion{H}{1} disk is embedded in a massive and extended triaxial dark
halo with a rotating figure, forcing the bar and spiral pattern at a
certain frequency, or ii) (some) dark matter is distributed in a disk
closely following the \ion{H}{1} distribution, rendering the disk
gravitationally unstable. Although both solutions lack direct
observational support (but see \citealt*{pcm94} for the latter), they
offer a way forward.

The purpose of this paper is to investigate whether either (or both)
of these mechanisms works in practice. The rotating halo shall be
modeled by external forcing from a rotating triaxial potential (which
could be due to a rotating triaxial halo, bulge, central bar, or a
mixture of those), and the massive disk will be modeled by simply
scaling up the gaseous surface density of the disk. We test the models
by running a simple two-dimensional (hereafter 2D) hydrodynamic code,
exploring the available parameter space, and comparing with the
observations. In \S~\ref{sec:method}, we first present the relevant
parameters for both sets of models and describe suitable observational
constraints. A brief description of the code is provided in
\S~\ref{sec:hydro_code}. The results for the external perturber models
are presented and discussed in \S~\ref{sec:triaxial_halos}, while
those for the massive disk models are described in
\S~\ref{sec:heavy_disks}. We summarize all results and conclude
briefly in \S~\ref{sec:conclusions}.
\section{Disk Response to a Rotating Non-Axisymmetric Forcing
Potential\label{sec:method}}
\nopagebreak
Fig.~\ref{fig:iv_images} shows the total \ion{H}{1} map and velocity
field of \objectname[]{NGC~2915} from the naturally-weighted cube of
\citet{mcbf96}, which we will use here for comparison purposes. The
high surface brightness core of \objectname[]{NGC~2915} (blue and
lumpy) is entirely contained within one beam width. Azimuthal averages
of the gaseous surface density $\Sigma_g\equiv4/3\,\Sigma_{HI}$, the
circular velocity $v_c$, and the \ion{H}{1} velocity dispersion
$\sigma_v$ are presented in Fig.~\ref{fig:ivs_profiles}
\citep[see][]{mcbf96,bfpm99}. These profiles are based on higher
spatial resolution uniformly-weighted data for $r\la10$~kpc and
naturally-weighted data at larger radii. The \ion{H}{1} has a central
peak but decreases slowly in the outer parts. The circular velocity
rises within $r\approx5$~kpc and levels out at
$v_c\approx80$~km~s$^{-1}$ at larger radii. The velocity dispersion is
the usual $8-10$~km~s$^{-1}$ outside 5~kpc but is much higher in the
central regions, justifying the use of an asymmetric drift correction
to the rotation curve by \citet{mcbf96}. We note however that the
velocity dispersion in the central regions may be overestimated due to
beam smearing effects (although \citealt{mcbf96} argue against this),
and that in turn the circular velocity may be
over-corrected (i.e.\ overestimated) within $r\la3$~kpc. The Toomre
$Q$ parameter \citep{t64} estimated from this data is high everywhere
\citep[$Q\ga6$; see][]{bfpm99}, implying that gravitational
instabilities such as SWING \citep{t81} can not grow if one assumes
that the observed gas surface density accounts for the entire disk
mass. As mentioned above, to force the emergence of spiral arms, one
can then study the response of the disk to an external forcing
potential or artificially increase the disk mass.

\placefigure{fig:iv_images}

\placefigure{fig:ivs_profiles}

To fit the models to the observations, one needs to identify the
relevant constraints provided by the data and the parameters one can
adjust to improve the match. The basic constraints are provided by the
first three moments of the data cube, shown in
Fig.~\ref{fig:iv_images}--\ref{fig:ivs_profiles}, but we refine these
below.
\subsection{Constraints from the Column Density Map\label{sec:method_i}}
\nopagebreak
Although the disk response in \objectname[]{NGC~2915} is likely
non-linear, it is instructive to look at the dispersion relation for
an $m$-folded density wave in a thin isothermal disk of sound speed
$c_s$, in the WKB limit \citep{ls64}:
\begin{equation}
\label{eq:disp_rel}
m^2(\Omega-\Omega_p)^2=\kappa^2-2\pi G\Sigma_g|k|+k^2c_s^2,
\end{equation}
where $k^2=m^2/r^2+k_r^2$, $k_r$ is the radial wave vector, $\Omega_p$
is the pattern frequency of the bar and spiral arms (i.e.\ the pattern
frequency of the forcing potential), $\Omega$ is the orbital
frequency ($v_c/r$), and $\kappa$ is the epicyclic frequency. 
We shall neglect
hereafter the possibility that the spiral arms have a different
(slower) pattern speed than the bar, see e.g.\ \citet{mt97} and
references therein. Inspection of the amplitude spectra (or
periodograms) of the $m=2$ perturbations which appear in our runs
shows that this assumption is reasonable. More precisely, one could
think of a halo with a slow figure rotation, the ILR of which would coincide with
the disk spiral corotation. We have performed early runs with halo pattern speeds
down to $\Omega_p=3$~km~s$^{-1}$~kpc$^{-1}$, and the power spectra of the disk response
show no trace at all of any response at any other frequency than $\Omega_p$. This
is not entirely a surprise however: on the contrary to what happens for a massive
collisionless disk, the spiral density waves propagate {\em inwards} from the ILR
and {\em outwards} from the OLR, while most of the [ILR, OLR] band is a forbidden
band for density waves propagation. The wave that a slow halo linearly excites at its
ILR (at large radius) can therefore naturally propagate inwards, and one does not
need to invoke non-linear coupling to account for a disk response in  the
disk. Furthermore, non-linear coupling efficiency is weakened in the present
situation with respect to a stellar disk. Indeed, non-linear coupling can be shown
to be efficient whenever the resonances of partner waves coincide (for instance corotation
and Lindblad resonance).
A simple argument for that is that the waves spend
a lot of time there, since their group velocities vanish, and they therefore have a 
large amount of time to exchange, non-linearly, a significant amount of energy and
angular momentum. In a collisionless system the integral $\int_r^{r_{LR}}dr/c_g$ diverges, where
$r_{LR}$ is a Lindblad resonance location and $c_g$ is the wave group velocity, which
makes non-linear coupling a very efficient process in stellar disks \citep{tsap87}. In a
non or weakly self-gravitating gaseous disk however, this integral is finite, and the non-linear coupling
efficiency is therefore much lower with respect to a collisionless situation.

We shall
assume hereafter that the dynamics of the \ion{H}{1} disk of
\objectname[]{NGC~2915} can be adequately modeled with an isothermal
gas disk in which the sound speed is equal everywhere to the observed
velocity dispersion, assumed to be isotropic:
$c_s\equiv\sigma_v$. Eq.~(\ref{eq:disp_rel}) shows that once
$\Omega_p$ is fixed, $k$ (and therefore $k_r$) follows, as $\Omega$,
$\kappa$, $\Sigma_g$, and $c_s$ are all observables. More precisely,
$k$ is given by a second-order polynomial equation admitting two
roots, corresponding to the so-called long and short waves. In the
regime where $Q$ is high (i.e.\ when the term $-2\pi G\Sigma_g|k|$ is
negligible), the two roots are very close to each other, and one can
consider only the short wave.

$k_r$ is related to the pitch-angle $\beta$ of the spiral pattern by
\begin{equation}
\label{eq:pitch}
\tan\beta=\frac{m}{k_rr}.
\end{equation}
Therefore, given the observations and a guess at the pattern frequency
of the perturber (and thus, in a forced regime, of the wave),
Eqs.~(\ref{eq:disp_rel}) and (\ref{eq:pitch}) directly yield the pitch
angle of the spiral pattern. One thus only needs to adjust the
perturber frequency to match the observed pitch angle, which can be
estimated observationally by $\tan\beta_{\rm obs}=\frac{d\log
r}{d\theta}$, where $r(\theta)$ traces the spiral arms and $\theta$ is
the azimuthal angle in a deprojected map of the galaxy. The
deprojected \ion{H}{1} surface density of \objectname[]{NGC~2915} in
$(\log r,\theta)$ coordinates is shown in
Fig.~\ref{fig:i_logrtheta_obs} for an inclination $i=53\fdg9$ (see
below). From this image, we obtain $\beta_{\rm obs}\approx7-10\degr$
for the main feature at $2.3\la\theta\la4$ (mod.\ $\pi$),
$r\approx12$~kpc, and $\beta_{\rm obs}\ga20\degr$ for the secondary
features such as $1.8\la\theta\la2.6$ (mod.\ $\pi$), $6\la
r\la11$~kpc.

\placefigure{fig:i_logrtheta_obs}

In the linear regime, the wave amplitude (which does not appear in the
dispersion relation) is independent of its pitch-angle. One can thus
successively adjust the perturber pattern frequency and strength in
order to match, respectively, the spiral pattern pitch-angle and
arm-interarm contrast. In \objectname[]{NGC~2915}, however, the
arm-interarm contrast is comparable to unity, the disk response is
likely non-linear, and one cannot fit the perturber parameters
($\Omega_p$ and strength) by simply exploring the wave pitch-angle and
arm-interarm contrast independently. The fit thus needs to be
undertaken through numerical simulations.

The constraints discussed above can be used to estimate the best set
of parameters for an external tidal perturber. In the second part of
this paper, in which we investigate the behavior of a $Q\approx1$
disk, the parameter space to be explored is different. It should also
be noted that most of the torque from the perturbing potential is
received by the disk at the Inner and Outer Lindblad resonances, where
$\Omega=\Omega_p\pm\frac{\kappa}{m}$. The radial profile of the
perturbing potential should thus only have a minor impact on the disk
response. This greatly simplifies the exploration of parameter space,
as it is not required to have a precise prescription of the
perturber's shape and potential (unavailable anyway), and its strength
can be adjusted simply by an adequate scaling (e.g.\ scaling the mass
of a bar-like perturber or the triaxiality of a triaxial
halo). Unfortunately, this simultaneously prevents any attempt to
reach a detailed knowledge of the perturber's structure.
\subsection{Constraints from the Velocity Field\label{sec:method_v}}
\nopagebreak
For an infinitely thin disk, the observed (i.e.\ line-of-sight)
velocity field of the gas flow can be expressed as:
\begin{equation}
\label{eq:vrad}
v=[-u\cos(\theta-\varphi)+v\sin(\theta-\varphi)]\sin i,
\end{equation}
where $u$ and $v$ are the radial and azimuthal components of the
velocity in the galactocentric rest frame, $i$ is the inclination, and
$\theta$ and $\varphi$ are, respectively, the azimuth of the fluid
element under consideration and of the observer. The notation is
illustrated in Fig.~\ref{fig:notation}. Although the origin of azimuth
can be chosen arbitrarily, it is traditionally chosen along the major
axis of the galactic disk, on the receding side. For this choice of
origin, $\varphi=-\pi/2$. As the velocity field associated with an
$m$-folded density wave has an intrinsic $e^{im\theta}$ dependence in
azimuth, one can easily see from Eq.~(\ref{eq:vrad}) that it will lead
to $m-1$ and $m+1$ folded signatures in the deprojected perturbed
velocity field. Thus, the perturbed velocity field associated with an
$m=2$ spiral will have $m=1$ and $m=3$ signatures when deprojected,
and these can in principle be used to constrain the models.

\citet{c93} showed that the weights of the $m=3$ to $m=1$ deprojected
components increase steeply across corotation. This is true if the
$m=1$ component of the observed velocity field comes only from the
$m=2$ spiral pattern. However, in the case of \objectname[]{NGC~2915},
which is likely warped, this property does not hold and, as we shall
see, we prefer to discard the $m=1$ information. More recently,
\citet{fetal01} used the observed velocity information to reconstruct
the velocity field of \objectname[]{NGC~3631}, and they underlined the
important role of the $m=1$ and $m=3$ coefficients.

\placefigure{fig:notation}
\subsubsection{Case of the $m=1$ component:\label{sec:method_m=1}}
The $m=1$ component of the deprojected velocity field can come from an
$m=2$ perturbation to the velocity, but also from the $m=0$ component,
i.e.\ the unperturbed circular rotation of the galactic disk. As soon
as the inclination of the disk varies with radius, it is not possible
to disentangle which fraction of the $m=1$ deprojected velocity comes
from an $m=2$ density wave and which fraction comes from a warp. This
corresponds to the infamous bias introduced in warped tilted ring
models by so-called streaming motions, in this case horizontal $m=2$
perturbed velocities. Since the disk of \objectname[]{NGC~2915} may
be warped \citep{mcbf96}, the $m=1$ component of the
deprojected velocity field is not a suitable constraint for the
simulations, and we will rather focus on the $m=3$ component.
We thus neglect the possible warp of \objectname[]{NGC~2915}, but this
not expected to alter significantly our conclusions, either
qualitatively or quantitatively.

\subsubsection{Case of the $m=3$ component:\label{sec:method_m=3}}
The $m=3$ component of the deprojected velocity field can come from an
$m=2$ perturbation to the velocity, such as an $m=2$ spiral density
wave, as well as from an $m=4$ wave (which has $m=3$ and $m=5$
signatures). It can also arise from a wrong evaluation of the
inclination \citep[see][ and references therein]{t02}. An error $\Delta
i$ on the inclination will lead to a $O(\Delta i)$ $m=1$ residual, but
also to a weak $O(\Delta i^2)$ $m=3$ residual. Although the $m=3$
residual coming from an inclination error is weak ($\sim v_c\,\Delta
i^2$), it is important to compare it with the $m=3$ residual coming
from the bar or spiral perturbation, which is also
small. Fig.~\ref{fig:m=3_raverage} shows the radius-averaged value of
this component as a function of the assumed inclination of the disk of
\objectname[]{NGC~2915} (\citealt{bfpm99} used $56\pm3\degr$). The
solid line represents a second order polynomial fit to the data. One
can easily check that the fit scales as $v_c\,\Delta i^2$, as
expected.

\placefigure{fig:m=3_raverage}

Fig.~\ref{fig:m=3_rprofiles} shows the amplitude of the $m=3$
component of the deprojected velocity field (hereafter called
$V_3^{\rm obs}$) as a function of radius, for three special values of
the assumed inclination: 55\fdg0 (corresponding to the lowest data
point in Fig.~\ref{fig:m=3_raverage}), 53\fdg9 (corresponding to the
fit minimum), and 52\fdg7 (corresponding to the shape of the outermost
isodensity contours, assuming circularity). These profiles do not
markedly differ from each other, showing that the small uncertainty on
the disk inclination has little impact on $V_3^{\rm obs}(r)$. A warp
could affect this profile in a similar fashion to inclination
uncertainty, but again the impact on the profile's shape should be
minimal. We thus consider $V_3^{\rm obs}(r)$ an appropriate constraint
for the simulations, where the equivalent profile $V_3^{\rm sim}(r)$
is easily constructed.

\placefigure{fig:m=3_rprofiles}

The $V_3^{\rm obs}(r)$ constraint is implemented as follows: we define
the scalar product between two radial functions $f$ and $g$ as
\begin{equation}
\langle f|g\rangle=\int_0^{r_{\rm ext}}rf(r)g(r)dr,
\label{eq:scalar_product}
\end{equation}
and then search for the maximum of the function 
\begin{equation}
\chi=\frac{\langle V_3^{\rm obs}(r)|V_3^{\rm sim}(r)\rangle}
{[\langle V_3^{\rm obs}(r)|V_3^{\rm obs}(r)\rangle 
\langle V_3^{\rm sim}(r)|V_3^{\rm sim}(r)\rangle ]^\frac12}
\label{eq:chi}
\end{equation}
in both time $t$ and observer azimuth $\varphi$ ($V_3(r)$ depends on
$\varphi$), while keeping the inclination fixed at $53\fdg9$. By
construction, $|\chi|$ is always smaller than one, and it is exactly
one if $V_3^{\rm obs}(r)$ and $V_3^{\rm sim}(r)$ are proportional
(Schwarz's theorem). The closer $|\chi|$ is to unity, the more similar
$V_3^{\rm obs}(r)$ and $V_3^{\rm sim}(r)$ are. In the linear regime,
$\chi$ is independent of the amplitude of the perturbing potential,
which can thus be optimized in a second step, matching $V_3^{\rm
sim}(r)$ to $V_3^{\rm obs}(r)$ in both shape and absolute value.

It must be noted that the perturber's amplitude can be adjusted either
by matching the amplitude of the perturbed velocity (i.e.\ $V_3^{\rm
sim}(r)$) or by matching the amplitude of the perturbed density (i.e.\
the arm-interarm contrast). As indicated above, we adopted the first
procedure. Naturally, the disk response should also match the perturbed
density amplitude or the model must be discarded. The ratio of the
perturbed velocity amplitude to the perturbed density amplitude depends
not only on the wave frequency but also on the disk parameters. Given the
simplifying assumptions made, we do not require a perfect simultaneous
match, but simply that the arm-interarm contrast 
resembles the one observed.
We also note that the $r$-weighting in Eq.~(\ref{eq:scalar_product})
corresponds to a uniform weighting of the 2D deprojected map, and that
the deprojected beam is elongated parallel to the minor axis at all
points. The effect of beam smearing's resulting $m=2$ modulation on
$V_3^{\rm obs}(r)$ is not taken into account by our analysis, but it
is likely to be small compared to the $m=3$ component we are
interested in. The integral in Eq.~(\ref{eq:scalar_product}) should
also have a lower limit at approximately one beam width. In practice,
$r$-weighting and beam smearing tend to quash any $m\ne0$ component at
the center and thus make the integrand negligible for
$0<r\la45\arcsec$. One can therefore adopt either choice of lower
limit.
\subsection{Constraints from the Velocity Dispersion
Field\label{sec:method_s}}
\nopagebreak
We have now outlined how to take into account the constraints arising
from the zeroth and first moments of the \ion{H}{1} data cube (i.e.\
the column density and velocity fields). The second moment (i.e.\ the
velocity dispersion field) is directly imposed to the model as the
(azimuthally averaged) sound speed profile. We have also performed
runs with a uniform sound speed $c_s\equiv8$~km~s$^{-1}$, as often
observed in the outer parts of dwarfs, because the large central
velocity dispersion could be partly due to beam smearing.
\section{Hydrodynamic Code\label{sec:hydro_code}}
\nopagebreak
The hydrodynamic code used is a simple 2D Eulerian finite difference
program on a polar mesh
written from scratch by one of us,
and formerly used in another astrophysical context (protoplanet--protoplanetary
disk tidal interactions). 
The mesh is staggered, that is to say 
the vertically integrated pressure
and the surface density are zone centered, while the radial (resp.\ azimuthal)
velocity is defined at the interface between two radially (resp.\ azimuthally)
neighboring zones. The code solves discrete versions of the continuity and Euler equations,
the different terms of which are treated sequentially through an operator splitting
technique in much the same way as is done in ZEUS \citep{sn92}. 
The algorithms used
enforce mass and angular momentum conservation (for an isolated system) up to the computer accuracy.
The advection is based on a second order
upstream difference \citep{vl77}. 
There is no energy equation as the closure
relation is provided by the observed velocity dispersion, treated here
as the sound speed. 
This ensures that at any instant the
simulated disk has a sound speed coinciding with the observed velocity
dispersion. The drawback is that the disk quickly fragments into
clumps at low $Q$. This does not happen in our externally forced
disks, however, as the surface density is low. For our heavy disk
models, where the gas surface density is scaled up, a work-around is
discussed in \S~\ref{sec:heavy_disks}. As mentioned above, since beam
smearing can result in an overestimated dispersion near the center,
several velocity dispersion profiles have been tried with central
values lower than observed. 
They all gave very similar results, except for the disk response near the center,
which displays a relatively smooth but complex behavior at large
$\sigma_v$ and relatively strong shocks and a higher accretion rate onto the
center for $\sigma_v\equiv 8$~km~s$^{-1}$. These differences are of negligible importance,
however, both in term of the $\chi$ value (our scalar product gives
a small weight to the central parts by construction) or in terms of the beam convolved-synthetic maps.
To relieve the Courant condition and
increase the time step, azimuthal advection is treated so as to get
rid of the average azimuthal velocity at each radius
\citep{m00a,m00b}. In our case, however, the gain provided is
relatively small since we have to deal with strong radial streaming at
the center, where the disk response is bar-like. As we use a 2D code,
we effectively neglect the galaxy warp.

For our external perturber models, the forcing is provided either by a
bar or by a triaxial halo. The bar models are characterized by the bar
pattern speed $\Omega_p$ and its surface density
\begin{equation}
\label{eq:barsd}
\Sigma(x,y)=\Sigma_b\left[\left(x^n+\frac{y^n}{q^n}\right)^{\frac 1n}\right],
\end{equation}
where $x$ (respectively $y$) is measured along the bar major
(respectively minor) axis, $q\equiv b/a\leq1$ is the bar axis ratio,
and $n$ is a dimensionless index characterizing the shape of the bar
isodensity contours. $\Sigma_b(r)$ is the bar density profile along
its major axis, given by
\begin{equation}
\Sigma_b(r)=\Sigma_0\exp(-r/h_b),
\end{equation}
where $\Sigma_0$ is the bar central surface density and $h_b$ its
scalelength. The bar potential is thus entirely determined by the set
of parameters $h_b$, $q$, $n$, and the total mass of the bar
\begin{equation}
M_{\rm bar} = \int\!\!\!\int\Sigma(x,y)dxdy,
\end{equation}
while its time behavior is determined by the pattern speed $\Omega_p$.
The exponential bar profile does not have a cut-off radius and the
quadrupole term arising from it has a power-law rather than an
exponential behavior. The bar elements located at $r\gg h_b$
nevertheless contribute negligibly to the bar potential itself, which
is dominated by the contribution of the inner parts. We reemphasize
that the detailed characteristics of the bar are unimportant for the
disk tidal response (see \S~\ref{sec:method_i}). The crude bar
description provided by Eq.~(\ref{eq:barsd}) is thus sufficient for
our purposes.

In addition to the axisymmetric part of the bar potential, another
$m=0$ potential is applied to the disk in all of the bar forcing runs,
so as to match the circular velocity curve. It can be interpreted as
due to a spherical halo (which can not drive a wake in the \ion{H}{1}
disk).

The triaxial halo models are characterized by the halo pattern speed
$\Omega_p$ and the ratio of the tangential to radial acceleration of
an embedded test particle. The tangential acceleration has an $m=2$
dependence in azimuth and the acceleration ratio was chosen either
constant in radius or with dependences similar to the ones shown in
Fig.~\ref{fig:acc_ratio_profiles}. These acceleration ratios were
obtained by computing the equatorial plane potential of ellipsoidal
halos with isosurface-density axis ratios $a:b:c$ and major axis
density profiles $\rho(r)\propto(1+(r/r_c)^2)^{-1}$. Varying $c/a$
from $0.5$ to $1.5$ only has a small impact on these curves. The halo
flattening is thus totally unconstrained by our models and we adopt
$c\equiv a$ for all halo models.

\placefigure{fig:acc_ratio_profiles}

The radial acceleration is derived from the observed circular
velocity. Namely, we take the following prescription for the halo
potential $\phi_{\rm halo}$:
\begin{equation}
\phi_{\rm halo}(r,\theta,t)=\phi_{m=0}(r)+\phi_{m=2}(r,\theta,t),
\end{equation}
where
\begin{equation}
\label{eq:phi_m=0}
\phi_{m=0}(r)=-\int_r^{+\infty}\frac{v_c(r')^2}{r'}dr'
\end{equation}
and
\begin{equation}
\phi_{m=2}(r,\theta,t)=\frac12{\cal R}{v_c(r)^2}\sin[2(\theta-\Omega_pt)],
\end{equation}
with
\begin{equation}
\label{eq:ratio}
{\cal R}=\left(0.34+\frac{0.155}{1+r/r_c}\right)\left(1-\frac ba\right)
\end{equation}
being the tangential to radial acceleration ratio ($r_c$ is the halo
core radius). Eq.~(\ref{eq:ratio}) provides a satisfactory fit to the
curves displayed in Fig.~\ref{fig:acc_ratio_profiles}.

Self-gravity is taken into account in all simulations but the stellar
content of \objectname[]{NGC~2915} is neglected. The disk potential is
evaluated through fast Fourier transforms (FFTs; see
\citealt{bt87}). Although potential evaluation with this technique is
affected by numerical artifacts, the effect is unimportant in our case
because self-gravity is weak ($Q$ is large and the gas contribution to
the total potential small).

\section{External Perturber Models\label{sec:triaxial_halos}}
\nopagebreak
\subsection{Bar Forcing Runs}
\nopagebreak
Figs.~\ref{fig:barruns1}--\ref{fig:barruns3} show the best match
results for a series of bar forcing runs. The pattern frequencies
shown range from $5.0$ to $7.5$~km~s$^{-1}$~kpc$^{-1}$ and the resolution
was $N_r\times N_\theta=100\times250$ for all runs, with an inner grid
boundary $R_{\rm min}=300$~pc and an outer boundary $R_{\rm
max}=25000$~pc, further than the edge of the observed \ion{H}{1} disk
to avoid spurious edge effects. The runs were performed in two steps.
First, a tentative bar mass $M_b^0=5\times10^9$~\msun\ was
adopted. The best match (both in time and observer's azimuthal angle
$\varphi$; see \S~\ref{sec:method_m=3} for details) was then
identified and the scaling ratio
\begin{equation}
\label{eq:scaling}
{\cal S} = \frac{\int_0^{r_{\rm ext}}rV_3^{\rm sim}dr}{\int_0^{r_{\rm ext}}rV_3^{\rm obs}dr}
\end{equation}
derived. A second run with bar mass $M_b=M_b^0/{\cal S}$ was then
performed to match the $m=3$ velocity profile in both shape and
amplitude. The bar masses used for this second step are indicated in
Tab.~\ref{tab:barmasses}. The adopted axis ratio $q=0.5$ and isophote
shape index $n=3$ are chosen from typical values found in the
literature and do not represent extreme cases
\citep{m95,f96}. Fig.~\ref{fig:barruns1} shows a comparison of the
$V_3^{\rm obs}$ and $V_3^{\rm sim}$ radial profiles for the best match
of each run. Synthetic maps of the corresponding \ion{H}{1} column
density and line-of-sight velocity where also produced and are shown
in Figs.~\ref{fig:barruns2} and~\ref{fig:barruns3}, respectively. 
We note that since the external perturber has central symmetry, and
since $N_\theta$ is an even number, the simulations can be run either
on a full or half grid. We have adopted the latter. This is not a
problem, however, as the simulations are not demanding on modern, cheap
platforms, and the flow remains symmetric.

\placetable{tab:barmasses}

\placefigure{fig:barruns1}

\placefigure{fig:barruns2}

\placefigure{fig:barruns3}

Contrary to our standard procedure (see \S~\ref{sec:method_s}), the
velocity dispersion for these runs was set to $8$~km~s$^{-1}$
everywhere. As discussed above, a first series of runs using the observed velocity
dispersion profile showed a non bar-like response and complex spiral
structures in the center. When lowering the sound speed in
the inner parts, one recovers the classical gas response in a bar-like
potential, as shown in
Figs.~\ref{fig:barruns1}--\ref{fig:barruns3}. The observed rise of the
velocity dispersion in the central part of the galaxy could then be
assigned to beam smearing, as the beam there intercepts kinematically
independent regions, but this is rather unlikely (see \S~4.2 and 5.2
in \citealt{mcbf96}). The high central dispersion is
almost certainly due to the burst of star formation. It may thus
be that the central gas dynamics cannot be properly modeled with
an isothermal gas, the sound speed of which corresponds to the
observed velocity dispersion, or simply that the gas dynamics near the
center is not dominated by the tidal forcing of a relatively slow,
large-scale perturber.
The initial setup is that for an unperturbed axisymmetric disk.
Contrary to \citet{a92b}, the external perturber is not turned on
slowly and a transient behavior is observed for the first few
dynamical times.

We note that in the synthetic maps, the bar position angle is {\em
not} adjusted; it is imposed by the best match observer's azimuth
$\varphi$ yielded by the $V_3^{\rm obs}(r)$ fitting (see Eq.~\ref{eq:chi}). The fact that the
modeled and observed bar position angles should coincide provides an
additional constraint to the models. Figs.~\ref{fig:barruns1}
and~\ref{fig:barruns2} show that the pattern speeds matching the
$V_3^{\rm obs}$ profile and the column density map (with a
satisfactory position angle for the bar) are relatively well
constrained, $\Omega_p\approx5.5-6.5$~km~s$^{-1}$~kpc$^{-1}$. For
these forcing frequencies, however, the isovelocity contours north of
the center exhibit a twist that is not observed \citep[see
Fig.~\ref{fig:iv_images};][]{mcbf96}.

The difference between the observed and synthetic velocities could be
interpreted as a kinematic warp, although its amplitude should be
large to account for such a twist. The disk response in the central
parts should not be too much of a concern, however, because of our
poor knowledge of the true gas velocity dispersion there, as well as
beam smearing effects on both the velocity and velocity dispersion
maps. The frequency range
$\Omega_p\approx5.5-6.5$~km~s$^{-1}$~kpc$^{-1}$ yields the right bar
position angle and a satisfactory response in the outer disk, both in
terms of perturbed surface density and streaming motions. However, the
main spiral feature observed in Fig.~\ref{fig:i_logrtheta_obs} is not
present in this series of runs. Fig.~\ref{fig:logrtheta_barun} shows
the disk surface density in $(\log r,\theta)$ coordinates for the best
bar forcing run, with $\Omega_p=6.0$~km~s$^{-1}$~kpc$^{-1}$. The main
characteristics of the observed distribution are recovered (end of the
bar around $3-4$~kpc, pitch angle $\approx20\degr$ for the secondary
outer features), but only a very faint structure with a low
pitch-angle is present in the outer disk ($r\ga10$~kpc). To be fair,
the observed \ion{H}{1} disk of NGC~2915 is extremely clumpy and
does not have a strict central symmetry. The bar-like structure is also
clearly asymmetric and the pitch-angle estimate for the southern outer
arm depends strongly on the \ion{H}{1} clump at about
$(-325\arcsec,+100\arcsec)$ in Fig.~\ref{fig:iv_images}.
Nevertheless, the residual velocity map does show mainly $m=1$ and $m=3$
components, strongly suggesting that, despite the departure of the
\ion{H}{1} column density map from central symmetry, the disk mainly owes
its structure to a grand design $m=2$ spiral wave.

\placefigure{fig:logrtheta_barun}

Bar forcing runs have also been performed with higher and lower
forcing frequencies than the ones presented here. For lower
frequencies, the bar-like response of the disk becomes naturally
longer and spreads over the whole disk. At higher frequencies, the bar
becomes increasingly short and eventually vanishes, while the spiral
response in the outer disk fades away. No value of the correlation
coefficient $\chi$ close to unity is ever reached in these low and
high frequency runs, so the perturber frequency is very well
constrained. We also tried a series of runs in which the bar
scalelength $h_b$ was varied (inversely proportional to the pattern
frequency), but no significant difference was found compared to the
results already presented. Finally, runs with different $q$ and $n$
were done, again without fundamental differences to the set
presented. The parameters $M_b$, $q$, $n$, and $h_b$ are degenerate,
i.e.\ different sets of values for these parameters can lead to the
same quadrupole term at the bar Lindblad resonances, where the disk is
torqued. More precisely, for a smaller bar mass, runs with slightly
more extreme values of $q$ and $n$ yield a similar disk response at
each forcing frequency.

The bars inferred from these runs must be dark. Indeed, the maximal
mass one can assign to the optical component of
\objectname[]{NGC~2915} is $M_*\approx7.4\times10^8$~\msun\ and the
\ion{H}{1} mass in the central bar is about $5\times10^7$~\msun\
\citep{mcbf96}, to be compared to the bar masses derived in our runs
of order $6\times10^9$~\msun\ (Tab.~\ref{tab:barmasses}). This stellar
mass is calculated using the circular velocity curve and the
exponential fit to the optical image from \citet{mmc94}, avoiding
contamination by the bright, young central objects. The optical image
further suggests that the stellar mass distribution is less elongated
than the bars used in our runs, although both are roughly
aligned. Increasing the bar axis ratio $b/a$ while preserving the fit would
require and even larger bar mass. The luminous component of
\objectname[]{NGC~2915} is thus totally unable to account for both the
observed arm-interarm contrast and the streaming motions amplitude in
the \ion{H}{1} disk.

In fact, given the maximum contribution of the stars
($\la20$~km~s$^{-1}$ outside 2-3~kpc) and the gas
($\approx15$~km~s$^{-1}$) to the circular velocity curve
\citep{mcbf96}, it is not surprising that they are unable to produce
non-circular motions of order $10$~km~s$^{-1}$ (as required). In an
unpublished manuscript, \citet{q98} estimates the non-circular motions
due to the gaseous bar at $2$~km~s$^{-1}$ at most, much smaller than
the observed \ion{H}{1} velocity dispersion. The gas response should
thus be smooth, unless a more massive non-axisymmetric component is
present.
\subsection{Triaxial Halo Runs}
\nopagebreak
We have also performed a series of runs in which the disk response is
excited by a rotating triaxial halo, as described in
\S~\ref{sec:hydro_code}, for the same set of pattern frequencies as
the bar forcing runs discussed above. The runs were again performed in
a two step procedure, first with an axis ratio $(b/a)_0=0.9$, then
with
\begin{equation}
\frac ba=1-\frac1{\cal S}\left[1-\left(\frac ba\right)_0\right].
\end{equation}
The axis ratios obtained for the second step vary between $0.83$ and
$0.88$ for pattern frequencies between $4.5$ and
$9.5$~km~s$^{-1}$~kpc$^{-1}$. The main characteristics of this set of
runs are comparable to the bar forcing runs, so we do not show all the
results explicitly. Fig.~\ref{fig:halores1} shows the surface density
response for a set of six different pattern speeds, at the best match
timestep for each case. Minor differences can be found compared to the
bar forcing runs, but the bar position angle and length are in
agreement and both sets display the same trends as the pattern speed
is varied. The highest $\chi$ values are obtained for roughly the same
forcing frequency range, but with slightly looser constraints
($6.5\la\Omega_p\la 8.0$~km~s$^{-1}$~kpc$^{-1}$). Note in Fig.~\ref{fig:halores1}
that $\chi$ remains high up to the last plot, corresponding
to $\Omega_p=8.5$~km~s$^{-1}$~kpc$^{-1}$, while it drops to lower values for 
the following forcing frequency ($\Omega_p=9.0$~km~s$^{-1}$~kpc$^{-1}$).
Both the position angle of the bar and the spiral pattern are
totally incompatible with observations for $\Omega_p=8.5$~km~s$^{-1}$~kpc$^{-1}$, however, which explains why we restrain our 
estimate of plausible values of $\Omega_p$ to $8$~km~s$^{-1}$~kpc$^{-1}$.

\placefigure{fig:halores1}
\subsection{Discussion of the External Perturber Models}
\nopagebreak
\subsubsection{Perturber Frequency}
\nopagebreak
We find that both our bar and halo forcing models can give rise to a
well-developed spiral pattern in a disk modeled on that of
\objectname[]{NGC~2915}. This is not a surprise, however, given that
external bar-like potentials have long been known to excite spiral
density waves in thin disks \citep[e.g.][]{gt79}. But it is interesting to
note that the halo triaxiality needed to account for the observed
non-circular motions is relatively mild, i.e.\
$b/a\approx0.83-0.88$. The pattern frequency range inferred from our
models is also consistent (within the errors) with the direct measurement
by \citeauthor{bfpm99} (\citeyear{bfpm99};
$\Omega_p=8.0\pm2.4$~km~s$^{-1}$~kpc$^{-1}$). The fact that the same
perturber strength can simultaneously account for the arm-interarm
contrast and the $V_3^{\rm obs}$ profile amplitude, and the fact that
the $V_3^{\rm obs}$ profile fit implies an observer's azimuth yielding
the correct bar position angle, both provide additional confidence in our models
and show that the detailed spiral structure of \objectname[]{NGC~2915}
can be explained by an external driver. Although it lies at the center
of the \ion{H}{1} bar, the stellar component of
\objectname[]{NGC~2915} is too small and lightweight to represent the
postulated bar-like perturber, and the spiral structure must be
excited by an unseen component.

Orbit calculations indicate that self-consistent bars should end
within their corotation radii \citep[e.g.][]{c80,a92a}. This imposes
in the case of \objectname[]{NGC~2915} an upper limit on the bar
pattern speed of $\Omega_p\la17.5$~km~s$^{-1}$~kpc$^{-1}$ \citep[for
$r_b\approx180\arcsec=4.6$~kpc; see][]{bfpm99}. Furthermore, the
general agreement from $N$-body simulations \citep[e.g.][]{se81,as86}
and observations (e.g.\ \citealt{a92b}; see \citealt{e96} for a
review) is that bars are fast, i.e.\ they end just inside corotation,
with $r_{\rm cr}/r_b\approx1.2$. This is clearly not the case here,
but \objectname[]{NGC~2915} is also of much later type than any of the
objects considered so far. Although it is marginally rejected by our
results, the \ion{H}{1} bar in \objectname[]{NGC~2915} could even end
within its own inner Lindblad resonance (ILR), for which there is
limited evidence in late-type bars \citep[e.g.][]{ce93,e96}. Our
result of $\Omega_p=6.0\pm0.5$~km~s$^{-1}$~kpc$^{-1}$ implies a slow
bar, with corotation near the edge of the observed \ion{H}{1} disk
($r_{\rm cr}/r_b\approx2.8$). This fact, together with
\objectname[]{NGC~2915}'s unusually dense and compact halo
\citep[see][]{mcbf96}, is consistent with models of dark matter
dominated (i.e.\ sub-maximal) barred galaxies, where bars transfer
most of their angular momenta to the halos
\citep[e.g.][]{w85,ds98}. Unfortunately, most of these works
considered collisionless (i.e.\ stellar only) systems, so it still
remains to be clarified exactly how relevant all these mechanisms are
in the case of \objectname[]{NGC~2915}.

Structure formation simulations in cold dark matter (CDM) scenarios
have long argued for (strongly) triaxial dark matter halos around
galaxies \citep[e.g.][]{fwde88,dc91,wqsz92}. These halos are supported
by anisotropic velocity dispersions, tend to be prolate (although both
the oblateness and triaxiality are slightly stronger in the outer
parts), and have angular momenta preferentially aligned with the minor
axis at all radii. Dissipation increases the oblateness but leaves the
flattening unchanged \citep{d94}. More recent COBE-normalized
$\Lambda$CDM simulations yield less flattened and less triaxial (i.e.\
more spherical) halos \citep{b02}. The same tendency is observed as a
function of decreasing mass and redshift. \citet{js02} also point out
that the high density (central) regions of halos in $\Lambda$CDM
universes are slightly more triaxial than the outer parts, but that
highly concentrated halos tend to be rather spherical. This would
argue against strong triaxiality in \objectname[]{NGC~2915}. The
probability distributions obtained by \citet{js02} argue against a
short-to-long axis ratio of $0.83-0.88$ (which is not ``flattened''
enough), but given this value a short-to-intermediate axis ratio of
$0.83-0.88$ is most likely. Indications are that halos in warm and
self-interacting dark matter simulations are even more spherical
\citep{b02,c03}. 
Observations of polar-rings, warps, and \ion{H}{1}
flaring suggest flattened halos around spiral galaxies
\citep[e.g.][]{srjf94,o96,bc97}. There is also evidence for weak
triaxiality \citep{fz92,absgw01}, but it is rather contradictory in
the case of elliptical galaxies (e.g.\ \citealt*{fgz94};
\citealt{bjcg02}).

As far as we know, the figure rotation of the triaxial halos formed in
CDM simulations has not been studied, beside the preliminary
investigation presented in \citet{bfpm99}. However, the forcing
pattern frequency inferred from our simulations is probably too high
to be accounted for by a rotating triaxial halo. This can be shown by
a simple calculation of the modified halo spin parameter $\lambda'$,
defined by \citet{bdkkkpp01} to be
\begin{equation}
\lambda'=\frac{J}{\sqrt{2}MVR},
\end{equation}
where $J$ is the angular momentum inside a sphere of radius $R$ (we
neglect in this order of magnitude calculation the small
triaxiality of the halo), containing a mass $M$, and $V=\sqrt{GM/R}$ is
the circular velocity imposed by the halo at radius
$R\,$. \citet{bdkkkpp01} find the log-normal statistics of the
$\lambda'$ parameter robust to the choice of $R$, so we choose here
the disk outer radius, $R\approx15$~kpc. The halo angular momentum up
to $R$ then reduces, summing on every dark particle~$i$ of mass $m_i$
and distance to the center $R_i$, to the well known expression
\begin{equation}
J=\frac23\Omega_p\sum_{i,R_i<R}m_iR_i^2,
\end{equation}
where we have assumed spherical symmetry, and where we made the
simplifying assumption that the motions of the dark particles in the
rotating frame do not contribute sizably to $J$, i.e.\ that there are
no net internal streaming motions in the rotating frame which could
sizably modify our estimate of the halo angular momentum. One can
thus write
\begin{equation}
\label{eq:lambdaprime_dmdr}
\lambda'=\frac{\sqrt{2}\,\Omega_p\int_0^R\frac{dM}{dr}r^2dr}{3MVR}.
\end{equation}

Integrating by parts, and noting that according to our estimate of
$\Omega_p$ the external perturber has its corotation at the edge of
the \ion{H}{1} disk (i.e.\ $\Omega_pR/V\approx1$), one obtains
\begin{equation}
\label{eq:lambda2}
\lambda'=\frac{\sqrt 2}{3}\left(1-2\int_0^1x^2\tilde v^2(x)dx\right),
\end{equation}
where $x=r/R$ and $\tilde v=v_c/V$. Noting that $\tilde v(1)=1$ and
$\tilde v(x)<1$ for $x<1$ (i.e.\ the circular velocity curve reaches
its maximum on $[0,R]$ at $r=R$), one obtains $1/3$ as a conservative
lower limit for the bracket in Eq.~(\ref{eq:lambda2}). We thus have
\begin{equation}
\lambda'>\lambda'_{\rm min}=\sqrt{2}/9\approx0.157.
\end{equation}

The statistics of $\lambda'$ is a log-normal distribution,
\begin{equation}
P(\lambda')=\frac{1}{\lambda'\sqrt{2\pi}\sigma}\exp\left(-\frac{\log^2(\lambda'/\lambda_0')}{2\sigma^2}\right),
\end{equation}
with $\lambda_0'=0.035\pm0.005$ and $\sigma=0.5\pm0.03$
\citep{bdkkkpp01}. If we adopt the most favorable values for our case,
$\lambda_0'=0.04$ and $\sigma=0.53$, the probability that a halo has a
spin parameter as large as $\lambda'_{\rm min}$ or higher is only
$5\times10^{-3}$. This probability drops to $1.6\times10^{-4}$ if one
takes the central values of $\lambda_0'=0.035$ and $\sigma=0.5$ and if
one uses the true behavior of $\tilde v(x)$ to estimate $\lambda'$
from Eq.~(\ref{eq:lambda2}), which leads to
$\lambda'\approx0.212$. Consequently, if the spiral pattern of
\objectname[]{NGC~2915} is driven by an external perturber, it is
unlikely that the perturber is a rotating triaxial halo, as this would
imply that a considerable fraction of the halo total energy is in
rotational kinetic energy ($\frac12 J\Omega_p$).

As stated, our estimate of $J$ relies on the absence of net internal
motions in the halo, which could sizably contribute to the halo
angular momentum (although this need not be the case). A similar issue
has been raised for stellar bars. \citet{wt83} and \citet{tw84b}
report that a bar can have a negative effective moment of inertia. In
our case, however, we need an almost exact cancellation of the solid
body estimate by internal motions to bring the spin parameter down to
plausible values, otherwise the order of magnitude of $\lambda'$ is
unchanged and our conclusions stand.

The alternative to the dark halo is a bar. Although unseen, this bar
should not be made of dark matter, as one would again face the problem
of excess angular momentum (see \citealt{cst95} and references
therein). On the other hand, if the statistics of the spin parameter
$\lambda$ were to change, triaxial halos with figure rotation should
then be expected to play a major role in galaxy formation and
evolution, allowing for example the driving of non-axisymmetric
structures in low surface brightness (LSB) galaxies.

Given that the total mass of \objectname[]{NGC~2915} is smaller than
most halos studied to date in CDM simulations \citep[${\cal
M}_T\ga2.7\times10^{10}$~\msun;][]{mcbf96}, it is of interest to study
the expected scaling of the triaxial figure pattern frequency with the
halo virial mass. An order of magnitude estimate for the former is:
\begin{equation}
\label{vir:1}
\Omega_{\rm vir}\sim\frac{V_{\rm vir}}{R_{\rm vir}},
\end{equation}
which corresponds to the pattern frequency of a bar with half-length
equal to the halo virial radius and $r_{\rm cr}/r_b=1$. It should thus
be considered the maximum figure pattern frequency for any halo with
extended triaxiality. From Eq.~(\ref{vir:1}), one can write
\begin{equation}
\label{vir:2}
\Omega_{\rm vir}^2\sim\frac{GM_{\rm vir}}{R_{\rm vir}^3}.
\end{equation}
The virial mass can be expressed as
\begin{equation}
\label{vir:3}
M_{\rm vir}=200\times\frac 43\pi R_{\rm vir}^3\rho_{\rm crit},
\end{equation}
where the factor $200$ depends only weakly upon the cosmology
\citep*[see e.g.][]{nfw96} and $\rho_{\rm crit}$ is the critical
density of the universe, linked to the Hubble constant $H$ by
\begin{equation}
\label{vir:4}
H^2=\frac{8\pi G}{3}\rho_{\rm crit}.
\end{equation}
Using Eqs.~(\ref{vir:2}) to~(\ref{vir:4}), one obtains
\begin{equation}
\label{vir:5}
\Omega_{\rm vir}\sim10H.
\end{equation}

This result is independent of $M_{\rm vir}$, so that triaxial figure
pattern frequencies should have statistics independent of the halo
virial mass. Assuming $H=70$~km~s$^{-1}$~Mpc$^{-1}$, one is led to
$\Omega_{\rm vir}\sim0.7$~km~s$^{-1}$~kpc$^{-1}$. The corresponding
figure rotation time is $t_{\rm fig}=2\pi/\Omega_{\rm vir}\sim9$~Gyr.
Interestingly, and although the details of the simulations are not
mentioned, this frequency is in good agreement with the preliminary
work presented by \citeauthor{bfpm99} (\citeyear{bfpm99}; $t_{\rm
fig}\approx5.5$~Gyr, see in particular their Fig.~10). The fact that
this pattern frequency is an order of magnitude lower than that
required to account for the spiral structure of
\objectname[]{NGC~2915} (if the latter is indeed driven by an external
perturber) strengthens our conclusions that external forcing is
unlikely.
\subsubsection{Generic Models}
\citet{bf02} recently published a study of gaseous disks immersed in
rotating triaxial halos modeled roughly on the situation in
\objectname[]{NGC~2915}. However, they do not attempt to fit
\objectname[]{NGC~2915} in detail and only consider the gas surface
density, not its velocities. Beside differences in the codes, they use
a uniform sound speed $c_s\equiv4$~km~s$^{-1}$, a total gas mass an
order of magnitude smaller than that in \objectname[]{NGC~2915}, and
start with a uniform gas surface density. In that sense, their study
is complementary to ours. \citet{bf02} also find a strong dependence
of the emerging disk structure on the perturber's frequency
$\Omega_p$, and their preferred values are consistent with ours (and
thus also unlikely). Similarly, they find that the gas self-gravity
and the scalelength of the dark halo are unimportant. Although their
parameter space is too large to be well constrained (with five free
parameters), they use an axial ratio $b/a=0.8$, similar to the one we
obtain for our triaxial halo runs. \citet{bf02} also explored the
effects of tilting the gas disk away from the equatorial plane of the
halo. They find rich morphologies and kinematics which could be
related to, e.g., polar rings and warps. Their runs also illustrate
that tilting alone does not give rise to spiral arms, as is expected
from the opposite vertical parities of the tilted halo perturbed
potential and the spiral density wave, showing that triaxiality (with
figure rotation) is essential.
\section{Heavy Disk Models\label{sec:heavy_disks}}
\nopagebreak
The simulations performed in the previous section showed that the
observations of \objectname[]{NGC~2915} could be reasonably well
accounted for by a grand-design spiral excited by an external
perturber, the corotation of which would lie close to the edge of the
\ion{H}{1} disk. The nature of this unseen perturber is problematic,
however, as a triaxial halo is unlikely to precess that fast, and the
nature of a hypothetic bar that would otherwise act as a driver is
unclear. Because the good agreements between the arm-interarm
contrast, the $m=3$ velocity perturbation amplitude and shape, and the
bar position angle strongly support the observed spiral structure
being a coherent grand-design pattern, and because the pattern speed
of this spiral is both observationally and numerically constrained to
have its corotation relatively close to the disk edge, it is tempting
to suggest a global disk instability as an alternative explanation for
the observed spiral structure. This requires the disk of
\objectname[]{NGC~2915} to be much more massive than what is inferred
from \ion{H}{1} column density measurements. As shown in
\citet{bfpm99}, \citet{t64,t81} parameters $X(r)=r\kappa^2/2\pi
m\mbox{G}\Sigma$ and $Q(r)=c_s\kappa/\pi\mbox{G}\Sigma$ are, for the
(luminous) disk of \objectname[]{NGC~2915}, everywhere well above
their critical values \citep[$X\approx3$ and $Q\approx2$ for a flat
rotation curve; see][]{t81}. The idea that (some) dark matter is
located in the disk is thus attractive since, in sufficient amount, it
can decrease $X$ and $Q$ below their critical values and render the
disk gravitationally unstable.

It is also well-documented that in many objects the \ion{H}{1}
surface density closely follows that of the total mass distribution.
In the outer parts of galaxies, where dark matter
dominates, \ion{H}{1} thus traces the surface density of dark matter (see
\citealt{b78,b81,b92a,b92b}; \citealt*{has01} for a recent
study). As pointed out by \citet{bfpm99}, this is also true
in \objectname[]{NGC~2915} of the additional disk material required to
bring $X$ and $Q$ to their critical values. The drawback is that by
lowering $Q$, one would also expect to see wide-scale star formation
in the disk, which is not observed \citep{bfpm99,mfbjka99}. This must
however depend on a number of parameters, including the nature of the
disk dark matter, and it is unclear whether \citeauthor{k89}'s
(\citeyear{k89}) star formation threshold can be applied directly to
\objectname[]{NGC~2915} (see \S~\ref{sec:discussion_disks} for more on
this issue). A number of independent lines of evidence also support
massive disks and/or light low-concentration halos (see, e.g.,
\citealt*{dmh96,dmh99} for the morphology of tidal
tails). Furthermore, \citet{f02} recently showed that massive disks
can account for the spiral structure observed in many LSB galaxies,
bypassing the need for an external perturber (but raising the issue of
the origin of such high ${\cal M}/L$ ratios). The support for disk
dark matter thus remains non-negligible.

\citeauthor{pcm94} (\citeyear{pcm94}; see also \citealt{pc94}) discuss
at length all the evidence in favor of rotationally supported (cold)
disk dark matter. The underlying idea of their work is that most of
the matter in disks is in the form of unseen cold molecular
hydrogen. The disks are close to the threshold of Jeans instability
and can develop some global disk gravitational spiral instability
endowing them with a spiral structure at large scale, while at small
scale they fragment into clumps inside which smaller scale fragmentation
occurs, recursively, until a cut-off scale is reached (the
``clumpuscule'' scale).

The numerical description of a fractally fragmented heavy disk
is an over-demanding problem ($Q\ga Q_{\rm crit}$, where $Q_{\rm crit}$ is
the critical value of the Toomre parameter below which the disk is
gravitationally unstable; $Q_{\rm crit}=1$ for an infinitely thin
disk). The spatial and temporal dynamic ranges required from the
simulations make it far beyond access of present computing resources,
and one can only address the problem partially. Here, we tackle the
problem of the global structure of a massive disk by starting with an
axisymmetric disk at equilibrium, to which a small $m=2$ seed is
added. The gravitational potential is then filtered out at each
timestep in order to retain only the large-scale $m=0$, $2$, $4$,
and~$6$ terms. We are thus able to follow a $Q\ga Q_{\rm crit}$ disk
large-scale dynamics over many rotation periods while avoiding the
formation of small-scale clumps, the study of which is not relevant to
our purposes. We filter out the $m=1$, $3$, and $5$ terms because an
initially axisymmetric system perturbed by a pure $m=2$ seed should
never display any odd-$m$ modes, even in the non-linear
stages. Indeed, as non-linear coupling of $m=2p$ and $m'=2q$ even
modes lead to energy and angular momentum transfer into $m_{\rm
nl}=m\pm m'=2(p\pm q)$ modes, any perturbation which arises in the
disk is always even folded (in other words, there is no spontaneous
breakdown of the central symmetry). In our situation, we ensure this odd-$m$
quiet behavior by filtering out odd modes at every time step.
\subsection{Heavy Disk Simulations and Parameter Space\label{sec:hdm}}
\nopagebreak
The heavy disk simulations use the same hydro-code as the external
perturber simulations (see \S~\ref{sec:hydro_code}
and~\ref{sec:triaxial_halos}), except for the aforementioned filtering
of the gravitational potential. Furthermore, the observed gaseous
surface density (Fig.~\ref{fig:ivs_profiles}) is scaled up by a
uniform scalar $\lambda_\Sigma\geq 1$. As the spiral structure driver
is now an instability internal to the disk, it is no longer necessary
to invoke any external triaxiality and the resulting disk is embedded
in a spherical halo. This halo radial density profile is constructed
for each $\lambda_\Sigma$ to match the observed circular velocity. We
used the observed velocity dispersion profile
(Fig.~\ref{fig:ivs_profiles}) directly and disregarded any possible
beam smearing or other effects. The only free parameter is thus the
dimensionless scalar $\lambda_\Sigma$, which we vary between $1$ and
$6$, at which value Toomre's $Q$ parameter is approximately unity at
$r=6.5$ and $9$~kpc and the disk becomes globally unstable to
axisymmetric modes \citep[see][]{bfpm99}. Our goal is to find the
range of $\lambda_\Sigma$ where a spiral instability develops, leading
to the formation of a grand-design spiral structure similar to the one
observed, and, if possible, to further constrain the value of
$\lambda_\Sigma$ so that the amplitude of the spiral also matches the
observed one. We again use the observed $m=3$ amplitude component of
the deprojected velocity field as our main constraint, in a four step
procedure:
\begin{enumerate}
\item For a given $\lambda_\Sigma$, we evaluate at each instant $t$
the observer's azimuth $\varphi$ that best fits the observed $V_3^{\rm
obs}$ profile, i.e.\ we maximize the correlation coefficient $\chi$
defined by Eq.~(\ref{eq:chi}) for each $t$.
\item Given $\varphi$, we calculate the relative perturbation
amplitude ${\cal S}(t)$ for each $t$, following
Eq.~(\ref{eq:scaling}).
\item We then compute the time average $\overline{\cal
S}(\lambda_\Sigma)$ of ${\cal S}(t)$, discarding the first $750$~Myr
(a few dynamical times) when transient behavior is observed (see
Fig.~\ref{fig:scaling}).
\item As $\overline{\cal S}(\lambda_\Sigma)$ should be unity in
order to match the observed spiral structure in amplitude, we vary
$\lambda_\Sigma$ and repeat the procedure until a satisfactory value
is found.
\end{enumerate}
\subsection{Heavy Disk Runs}
\nopagebreak
Fig.~\ref{fig:scaling} shows the relative amplitude of the spiral
instability obtained as a function of the scaling ratio
$\lambda_\Sigma$. $\overline{\cal S}(\lambda_\Sigma)$ depends
dramatically on the scaling, allowing us to constrain $\lambda_\Sigma$
accurately. The best match to the observed amplitude is obtained for
$\lambda_\Sigma=4.93$. Fig.~\ref{fig:bestheavy} shows the
corresponding density and velocity maps, as well as a comparison
between $V_3^{\rm obs}(r)$ and $V_3^{\rm sim}(r)$. We again stress
that the observer's azimuth in the three panels of
Fig.~\ref{fig:bestheavy} is the same, and that the position angle of
the spiral structure is {\em not} adjusted but is imposed by the best
match procedure. The corresponding $\chi$ value is
$0.982$. Fig.~\ref{fig:besthvlogrt} shows the surface density response
in $(\log r,\theta)$ coordinates. Although it is significantly
different from the observed one in the center (less bar-like), the
tightly wrapped spiral arms in the outer parts are well reproduced. In
particular, the pitch angle measured for these spiral arms is about
$7\degr$, consistent with the observations and in contrast to the
external forcing runs. Figs.~\ref{fig:scaling}--\ref{fig:besthvlogrt}
thus show that it is possible to account for most of the large-scale
structure in the disk of~\objectname[]{NGC~2915} by simply scaling up
uniformly the observed gas surface density.

\placefigure{fig:scaling}

\placefigure{fig:bestheavy}

\placefigure{fig:besthvlogrt}

Before inferring from this calculation the actual mass that has to be
in the disk of \objectname[]{NGC~2915}, if it indeed owes its spiral
structure to gravitational instabilities, it is important to estimate
the impact of finite thickness. The grid over which we compute the
potential has everywhere a zone size much smaller than the estimated
disk thickness. The computed potential is thus that of an infinitely
thin disk with surface density $\Sigma_{\rm thin}(r,\theta)$. This
potential, filtered out in $m$, can be written as
\begin{equation}
\label{eq:potentialthin}
\phi_{\rm thin}(r,\theta)=\phi_0^{\rm halo}(r)+\phi_0^{\rm disk}(r)+\sum_{n=1}^{3}\phi_{2n}^{\rm disk}(r,\theta),
\end{equation}
where $\phi_m^{\rm disk}(r,\theta)$ is the $m-$th Fourier component of
the infinitesimally thin disk's potential. On the other hand, for a
moderately thick (i.e.\ $H\ll r$, where $H$ is the characteristic
thickness) but vertically resolved disk with the same surface density,
the potential expression would read
\begin{equation}
\label{eq:potentialthick}
\phi_{\rm resolved}(r,\theta)=\phi_0^{\rm halo}(r)+\phi_0^{\rm disk}(r)+\sum_{n=1}^{3}q\,\phi_{2n}^{\rm disk}(r,\theta),
\end{equation}
where $q$ is a correction factor, the expression of which is
$q\approx(1+kH)^{-1}$ (\citealt{v70,r92}; $k$ is again the
wave-vector). Using Eq.~(\ref{eq:pitch}) and writing $k_r\approx k$
(since $m=2$ is small) yields
\begin{equation}
\label{eq:reduc}
q\approx\left(1+\frac{2\sigma_v}{v_c\tan\beta}\right)^{-1}.
\end{equation}
A thin but vertically resolved disk, leading to the same perturbed
potential as computed for our heavy disk runs, should therefore have
a surface density
\begin{equation}
\label{eq:resolveddisk}
\Sigma'_g(r)=\frac{\lambda_\Sigma}{q}\Sigma_g.
\end{equation}

In the outer disk of \objectname[]{NGC~2915}, assuming
$\sigma_v=8$~km~s$^{-1}$, $v_c=80$~km~s$^{-1}$, and $\beta=10\degr$,
$q=0.47$, while $q=0.65$ for $\beta=20\degr$. In the inner disk, the
wave becomes more open ($\tan\beta$ increases) but the circular
velocity decreases. We thus adopt for the sake of simplicity a uniform
value for our runs, $q=0.5$. This is obviously a crude assumption and
is the main source of uncertainty when computing the required disk surface
densities (much more than the small uncertainty on $\lambda_\Sigma$).
For such a disk, the fraction of the axisymmetric part of the
potential due to the disk is $q^{-1}\approx2$ larger than in our runs,
but since the halo is constructed afterward to fit the observed
circular velocity, the dynamics is not affected (the total
axisymmetric component of the potential $\phi_0(r)=\phi_0^{\rm
halo}(r)+\phi_0^{\rm disk}(r)$ remains unchanged). We can construct a
modified halo profile using the observed circular velocity curve, from
which we subtract the contribution of a disk with surface density
profile given by Eq.~(\ref{eq:resolveddisk}) (i.e.\ about ten times
more massive than the observed gaseous disk). The result for our best
heavy disk model with $\lambda_\Sigma=4.93$ is shown in
Fig.~\ref{fig:massfraction}.

\placefigure{fig:massfraction}

Fig.~\ref{fig:massfraction} also shows that the disk and halo circular
velocities, and thus their integrated masses, are roughly equal over
most of the disk. This result depends on the adopted value for the
correction factor but remains qualitatively unchanged as long as the
correction factor parameter $q\la\,0.65$. For $q=0.7$, the halo
dominates by a factor of $2$ at $r=10$~kpc and more further out. The
equality of disk and halo masses is in agreement with recent structure
formation simulations including gas cooling in standard $\Lambda$CDM
cosmologies (Teyssier, priv.\ com.; see also \citealt{te02}). Although
the stellar component of \objectname[]{NGC~2915} has been neglected in
all simulations, we do not expect any sizable consequences on our
result, as the curves in Fig.~\ref{fig:massfraction} would only need
to be modified in the inner few kiloparsecs. More precisely, as our
models respect the observed circular velocity profile, the optical
component of \objectname[]{NGC~2915} is effectively incorporated in
the halo.

We saw in Fig.~\ref{fig:bestheavy} that the spiral pattern obtained
differs somewhat from the observed one. In particular, the spiral arms
occur at a radius slightly smaller than observed and the central
condensation has a stronger $m=4$ symmetry. Nevertheless, the agreement between the
$\lambda_\Sigma=4.93$ model and the observations is satisfactory
considering the small dimensionality ($1$) of the parameter space. We
stress that any $m=4$ spiral density wave would have an effect on the
$m=3$ deprojected velocity field, so that the simulated $V_3^{\rm
sim}$ profile could arise from a mixture of $m=2$ and $m=4$ waves. In
this run, however, we find that the outer spiral structure is the
superposition of two $m=2$ spiral density waves with different pattern
speeds. The main wave has $\Omega_p=6.0$~km~s$^{-1}$~kpc$^{-1}$,
consistent with the direct measurement of \citet{bfpm99}, and
corresponding to a corotation close to the disk edge. The second wave
has $\Omega_p\approx10$~km~s$^{-1}$~kpc$^{-1}$ but is not well
determined, as the perturbed density power spectrum deduced from the
simulation is relatively noisy.
\subsection{Discussion of the Heavy Disk Models\label{sec:discussion_disks}}
\nopagebreak
Not surprisingly given our motivations for doing so, we find that
scaling up the gas surface density of \objectname[]{NGC~2915} gives
rise to a well-developed spiral pattern. Our heavy disk model also
provides a satisfactory fit to \objectname[]{NGC~2915}'s column
density map and velocity perturbations. The corrected scaling factor
$\lambda_\Sigma/q\approx10$ we obtain is in good agreement with those
derived for a variety of spiral galaxies, where typical values are
$6-10$ with a tail at larger values (e.g.\ \citealt{has01,c03};
$\lambda_\Sigma/q=\Sigma_{\rm dark}/\Sigma_g+1=3/4\,(\Sigma_{\rm
dark}/\Sigma_{HI})+1$). \citet{s99} obtains slightly lower factors for
his sample of (gas-rich) late-type dwarfs, but again a tail of high
values is present.

We note however that the above values are normally determined by
attempting to fit the circular velocity curve only (with a scaled-up
\ion{H}{1}+He disk and a stellar component with free ${\cal M}/L$),
while our goals here differ and we allow for an extra dark halo. Such
circular velocity curve fitting would yield a scaling factor of about
$25$ for \objectname[]{NGC~2915} (neglecting the stellar component,
which is safe here), but as can be seen from
Fig.~\ref{fig:massfraction} it would poorly reproduce the total circular
velocity. Wiggles in
the predicted circular velocity curve are not present in the observed
one, requiring a smoother dark disk material distribution for improved
fitting. Furthermore, in \objectname[]{NGC~2915}, the scaling factor
obtained from the circular velocity curve is much higher than that obtained
from the requirement of spiral structure sustainability, and the
resulting disk would be violently unstable to axisymmetric
perturbations (see below). This strongly argues for the presence of a
classical (i.e.\ spherical or near-spherical) dark matter halo in
addition to any realistic massive disk. A similar argument could be
made based on the number of spiral arms in the disk
\citep*[e.g.][]{abp87}. As pointed out by \citet{has01}, a circular
velocity curve remaining flat significantly outside of the observed
range would normally make the fit of a scaled-up gaseous disk alone
progressively more difficult, but this is not the case here as the
circular velocity curve predicted from the \ion{H}{1} is still
slightly rising in the outer parts.

The adopted scaling of the gas surface density brings the disk of
\objectname[]{NGC~2915} close to the critical value of the Toomre
parameter, for which the disk is unstable to axisymmetric modes and
fragments ($Q_{\rm crit}\approx0.5$ for a vertically resolved
disk). It is therefore interesting to compare the surface density
profile of \objectname[]{NGC~2915} with \citeauthor{k89}'s
(\citeyear{k89}) threshold for the onset of star formation, as the
disk is not observed to form stars. Kennicutt's critical surface
density is defined by
\begin{equation}
\label{eq:keni1}
\Sigma_c(r)=\alpha\frac{\kappa\sigma_v}{3.36 G},
\end{equation}
where $\alpha=0.67$. This critical surface density corresponds to the
marginal stability of a thin gas layer embedded in a more massive,
thicker stellar disk to two-component (fluid and collisionless)
axisymmetric instabilities (see \citealt{r01} and references
therein). If we adapt this criterion (marginal stability) to the
present situation, for which $Q(r)=\kappa\sigma_v/\pi G\Sigma$ and
$Q_{\rm crit}\approx0.5$, the critical surface density becomes
\begin{equation}
\label{eq:keni2}
\Sigma_c'(r)=\frac{2\kappa\sigma_v}{\pi G}\approx3.2\,\Sigma_c(r).
\end{equation}
We note that although our scaling adds another fluid to the disk of
\objectname[]{NGC~2915}, it is indistinguishable from the \ion{H}{1}
as the same sound speed is assumed, and we recover the one-fluid
stability criterion. Indeed \citep[see e.g.\ the $R=1$ dotted line in
Fig.~3 of][]{r01}, the stability criterion reads $1/Q_{\rm
disk\;DM}+1/Q_{\rm HI}=1/Q_{\rm crit}$ (where we replaced $1$ by
$1/Q_{\rm crit}$ to account for the vertical thickness), leading to
$\pi G(\Sigma_{\rm disk\;DM}+\Sigma_{\rm HI})/c_s\kappa=1/Q_{\rm
crit}$ and to Eq.~(\ref{eq:keni2}), where $\Sigma_c'(r)$ refers to the
{\em total} (i.e.\ dark matter + \ion{H}{1}) disk surface density.

The two critical surface densities given by Eqs.~(\ref{eq:keni1})
and~(\ref{eq:keni2}) are compared with that inferred from our best
match heavy disk model in Fig.~\ref{fig:keni}, where we use the
observed velocity dispersion profile $\sigma_v(r)$. According to the
standard criterion \citep[also used by][]{bfpm99}, the disk would be
expected to form stars at almost all radii, whereas according to the
modified criterion it is not expected to form stars at all. This
should not come as a surprise, however, since the modified criterion
comes from the same marginal stability argument underlying our massive
disk models, where we simply scaled up the observed gas surface
density, approaching the limit of gravitational instability without
ever reaching it. The correction factor $q=0.5$ used to infer the
surface density of a vertically resolved disk is the same as the
marginally (un)stable $Q$ value for a vertically resolved gaseous
disk, explaining why the heavy disk surface density profile is
everywhere close to but smaller than $\Sigma'_c(r)$. The
non-prediction of star formation within $r\approx3.5$~kpc, even with
the standard threshold, should not be taken too seriously since we
have neglected the stellar component of \objectname[]{NGC~2915}, which
extends to about 3~kpc \citep{mmc94}.

\placefigure{fig:keni}

It should be noted that the scaling factor $\lambda_\Sigma/q\approx10$
inferred from our heavy disk models depends on the assumed distance
$D$ to \objectname[]{NGC~2915}. Since the observed Toomre $Q$
parameter scales as $D^{-1}$, a smaller scaling ratio would be needed
if the galaxy were further than estimated by \citeauthor{mcbf96}
(\citeyear{mcbf96}; our adopted distance of $5.3\pm1.6$~Mpc). However,
an order of magnitude increase in the distance would be needed to
reach $Q(r)\approx0.5$, which seems highly unlikely. Nevertheless, an
improved distance determination would be helpful, for example from
Cepheids or the tip of the red giant branch with HST. \citet{q98}
discussed in detail the issue of \objectname[]{NGC~2915}'s
distance. While she considers an increase of the distance by a factor
of $2$ unlikely but possible, her arguments would completely reject an
increase by a factor of $10$.

We also stress that $Q$, and thus $\lambda_\Sigma$, also depend on the
assumed velocity dispersion of the additional disk material. For the
sake of simplicity and in order to keep our parameter space as small
as possible, we have assumed that the unseen material is tightly
linked to the observed \ion{H}{1} and has the same velocity dispersion
and (consistently) the same vertical extent. Less additional material
would be needed if it had a lower velocity dispersion. Indeed, the
ratio $c_s/\Sigma_g$ (or $\sigma_v/\Sigma_g$) is degenerate in the
expression of $Q$. A change of order unity in $Q_{crit}$ would also be
expected to account for two-component axisymmetric (in)stability, and
other two-components instabilities may develop \citep{r01}.
\section{Discussion and Conclusions\label{sec:conclusions}}
\nopagebreak
We have performed 2D hydrodynamic simulations of the galaxy
\objectname[]{NGC~2915} in order to find a plausible explanation for
the spiral structure observed in its extended \ion{H}{1} disk. We have
explored two main hypotheses: i) the observed bar/spiral structure is
excited by an external driver, which we assume to be either a rotating
bar or triaxial halo, and ii) the observed structure is due to a
gravitational instability in the disk, which would contain an
important unseen component and be much more massive than currently
inferred from \ion{H}{1} observations. In the first case, the free
parameters are the external driver pattern frequency and strength (bar
mass or halo triaxiality). Our results constrain the pattern frequency
to $\Omega_p\approx5.5-6.5$~km~s$^{-1}$~kpc$^{-1}$ and the bar mass
and halo axis ratio to $M_{\rm bar}\approx5-7\times10^9$~\msun\ and
$b/a\approx0.83-0.88$, respectively. However, based on current
structure formation simulations, it is unlikely that a triaxial halo
would have such a fast figure rotation (the spin parameter
$\lambda$ is too large), while the mass required for the bar is at least an order
of magnitude larger than the optical mass of the central
galaxy. All our external driver models also fail to reproduce the
small spiral arms pitch angle observed in the outer disk.
The external perturber model is thus implausible and we do not favor it. In the
heavy disk case, the only free parameter is the disk surface density
scaling ratio. We found that a (corrected) scaling ratio of about $10$
leads to a spiral instability adequately accounting for the disk
spiral structure and non-circular motions: the spiral
pitch angle and arm-interarm contrast are accurately reproduced. The scaling ratio is
well constrained by the simulations and the main uncertainty arises
from the correction factor of the perturbed potential (to account for
finite thickness). A scaling ratio of $10$ yields
roughly the same amount of disk matter and halo dark matter over the
disk radius while the halo dominates further out. This is consistent
with recent $\Lambda$CDM structure formation simulations.
Furthermore, given that the heavy disk model
reproduces the perturbed velocities and leads to a redistribution of
angular momentum, an external triaxial potential is not required to
explain the fueling of the central BCD starburst. Even for such a
heavy disk, \citeauthor{k89}'s (\citeyear{k89}) star formation
criterion is valid and accounts for the absence of star formation in
the disk of \objectname[]{NGC~2915}, provided it reads as $Q>Q_{\rm crit}$,
where $Q_{\rm crit}$ is the critical value of the Toomre parameter for
the onset of Jeans instability. As the disk of \objectname[]{NGC~2915}
does not lie in the equatorial plane of a more massive, thicker
stellar disk, the expression of $Q_{\rm crit}$ is different from that
of standard galactic gaseous disks.

Throughout this work, we have neglected a possible warp in
\objectname[]{NGC~2915}, we have assumed that the observed velocity
dispersion could be adequately modeled as the sound speed of an
isothermal gas, and we have not considered the small-scale clumpy
structure present at the beam scale in the \ion{H}{1} observations. We
have also assumed, respectively, a unique external driver pattern
frequency and a uniform disk surface density scaling ratio, in order
to keep the dimensionality of our parameter space as small as
possible. As such, our study is exploratory in nature. This might in
turn lead to too crude a description of \objectname[]{NGC~2915}, and
relaxing one or several of these assumptions may be required to
improve the fit. For our heavy disk models, we also assumed a
non-responsive spherically symmetric halo, whereas the disk mass is
comparable to the halo mass within its radius. Considering a live halo
thus appears a priority for any further similar modeling. It should in
particular improve the fit to the observed central bar-like structure,
where our current model fails.

Although we favor the heavy disk model over an external perturber,
both models fail to reproduce satisfactorily the gas response in the
center. Perturber models can loosely reproduce the bar-like feature
observed but are unable to account for the kinematic twists, while
heavy disk models simply fail to generate bar-like structures.
The complexity of the central dynamics is in any case probably much
greater than our models allow, and it is likely for instance that a
different forcing frequency has to be considered for the central
regions. In addition, the heavy disk models do not include any
interaction or feedback from the stars, which could initiate the
bar-like response. Despite a poor description of the central regions,
we emphasize that a satisfactory description of the outer disk
($4.6<r<15$~kpc) is obtained, simultaneously in terms of perturbed
velocity and perturbed density, with a very restricted set of
adjustable parameters ($2$ in the forced runs, $1$ for the heavy
disks). Furthermore, only the heavy disk models properly account for
the observed small pitch angle of the spiral arms. This strong
constraint thus favors this latter set of models.

If one adopts the massive disk interpretation, \objectname[]{NGC~2915}
appears as an aborted optical grand-design spiral because its gas
layer has not reached the critical surface density for star
formation. The spiral pattern is nevertheless present because the
surface density is {\em just} under this threshold. We recall however
that there is a degeneracy between the surface density and
velocity dispersion of the dark disk material. A long-standing and
possibly related issue is how the \ion{H}{1} velocity dispersion can
remain at $\approx8$~km~s$^{-1}$ in the outer disk, without any
obvious thermal or mechanical energy source. In particular, the
question whether the observed spiral density wave could constitute a
sufficient heating source for the disk material should be addressed
(whether due to an external driver or to a gravitationally unstable
massive disk).

It is interesting to note that if, in any isolated object, the
increased disk surface density required to explain the existence of a
spiral pattern were to over-predict the observed circular velocity
curve, then disk dark matter (exclusively) could be ruled out as the
source of that pattern, leaving a rotating triaxial halo as the most
likely source. Undisturbed LSB galaxies which are not dark matter
dominated and show a clear spiral structure would be ideal candidates
for such a test. However, it is not clear whether such objects
exist. Spiral-based scaling factors in LSB galaxies generally produce
maximum disk velocities that are high but still within the maximum
allowed by the circular velocity curve \citep[e.g.][]{f02}.

\acknowledgements Support for this work was provided by NASA through
Hubble Fellowship grant HST-HF-01136.01 awarded by the Space Telescope
Science Institute, which is operated by the Association of
Universities for Research in Astronomy, Inc., for NASA, under contract
NAS~5-26555. The authors wish to thank G.R.\ Meurer for
\objectname[]{NGC~2915}'s data and the organizers of the Guillermo
Haro Workshop held in Puebla, Mexico, in May~2001, which was the
starting point of this collaboration. F.M.\ wishes to thank R.\
Teyssier and P.-A.\ Duc for valuable discussions. M.B.\ similarly
acknowledges useful discussion with K.C.\ Freeman. J.\ van Gorkom,
F.\ Combes and E.\ Athanassoula are
also thanked for reading and commenting on a first draft of the
manuscript. The Digitized Sky Surveys were produced at the Space
Telescope Science Institute under U.S. Government grant NAG
W-2166. The images of these surveys are based on photographic data
obtained using the Oschin Schmidt Telescope on Palomar Mountain and
the UK Schmidt Telescope. The plates were processed into the present
compressed digital form with the permission of these
institutions. This research has made use of NASA's Astrophysics Data
System Bibliographic Services.

\clearpage
\figcaption[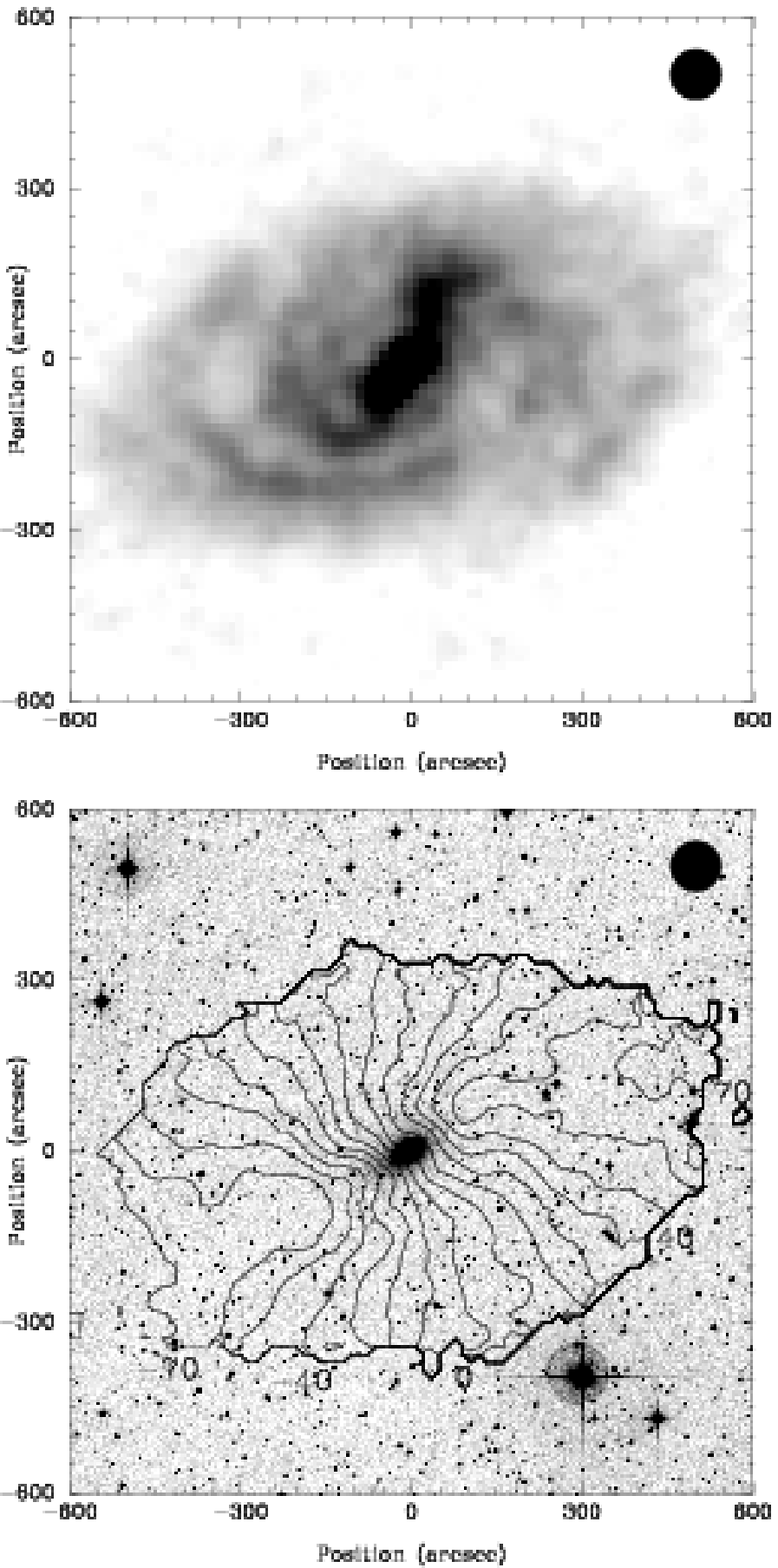]{First two moments of the \ion{H}{1} distribution
in NGC~2915 (natural weighting). {\em Top:} Grey scale image of the
total \ion{H}{1} map (moment~0). {\em Bottom:} Mean \ion{H}{1}
velocity field (moment~1) overlaid on an optical image from the
Digitized Sky Survey. The velocities are relative to the systemic
velocity of NGC~2915, the velocity resolution is 6.6~km~s$^{-1}$, and
the contours are spaced by 10~km~s$^{-1}$. The beam is
$45\arcsec\times45\arcsec$ in both maps and drawn to scale in the top
right corner of each panel. Adapted from \citet{mcbf96}
with permission.\label{fig:iv_images}}
\figcaption[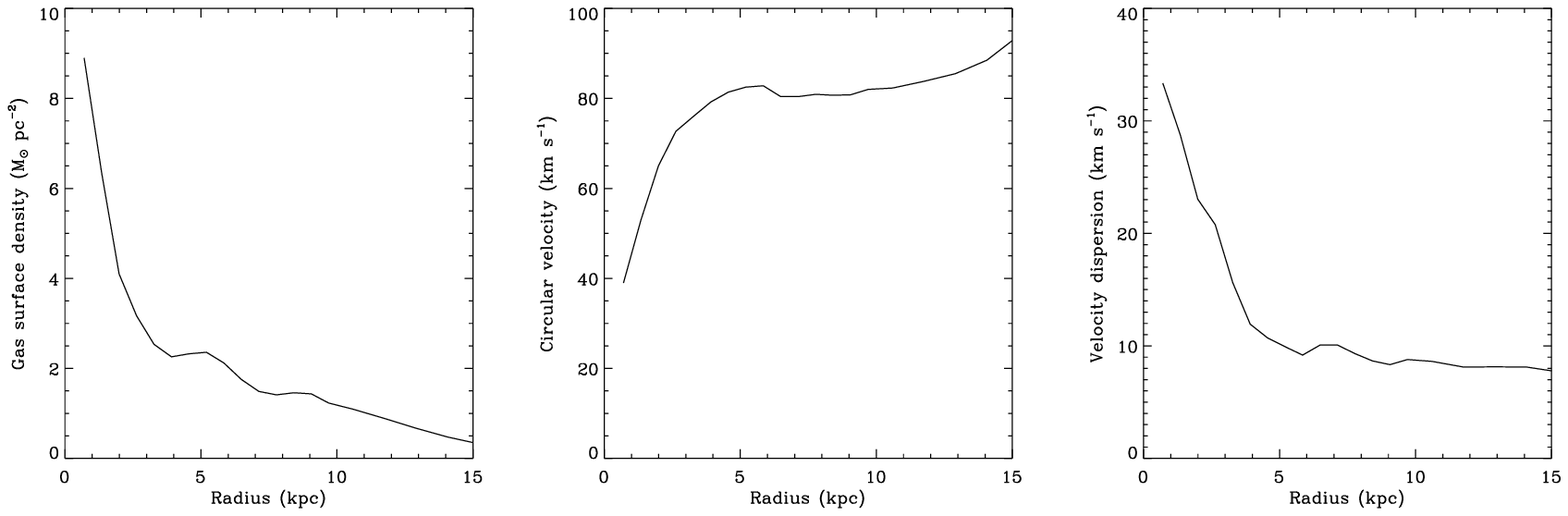]{Azimuthally averaged radial profiles of
NGC~2915. {\em Left:} Gas surface density $\Sigma_g$. {\em Center:}
Circular velocity $v_c$ (rotation curve corrected for asymmetric
drift). {\em Right:} \ion{H}{1} velocity dispersion $\sigma_v$. The
profiles are based on uniformly-weighted data (25\arcsec\ beam) for
$r\la10$~kpc and naturally-weighted data (45\arcsec\ beam) at larger
radii.\label{fig:ivs_profiles}}
\figcaption[fig3.eps]{Gas surface density of NGC~2915 in $(\log
r,\theta)$ coordinates, obtained by assuming an inclination
$i=53\fdg9$ and scaling the deprojected \ion{H}{1} surface density by
$4/3$.\label{fig:i_logrtheta_obs}}
\figcaption[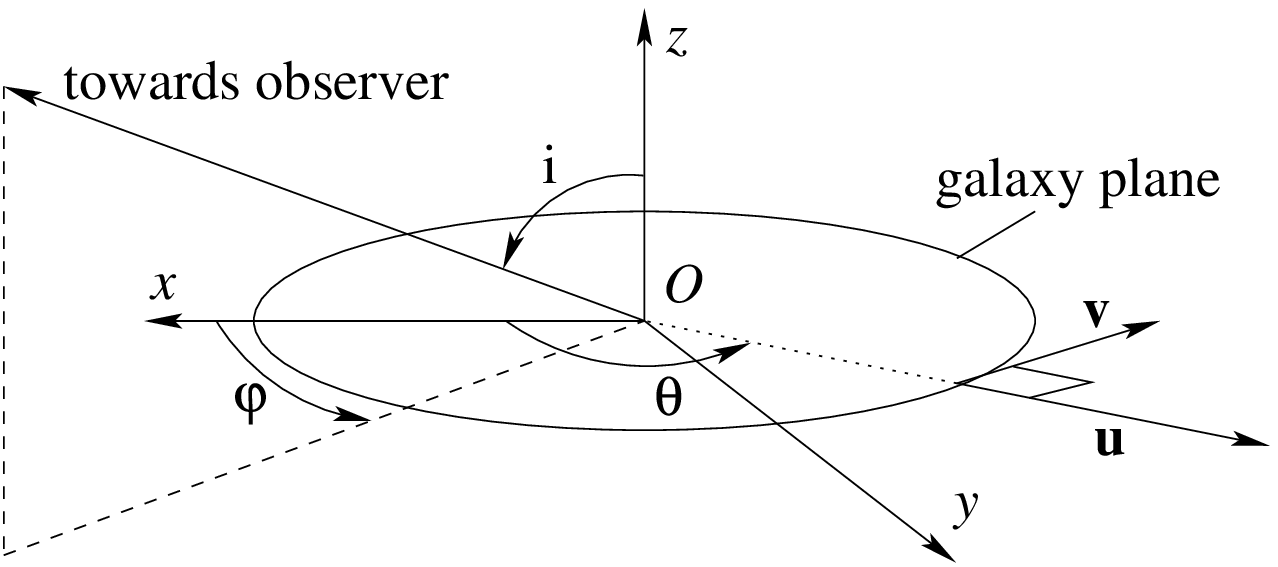]{Sketch of the notation used in the text
(\S~\ref{sec:method_v}).\label{fig:notation}}
\figcaption[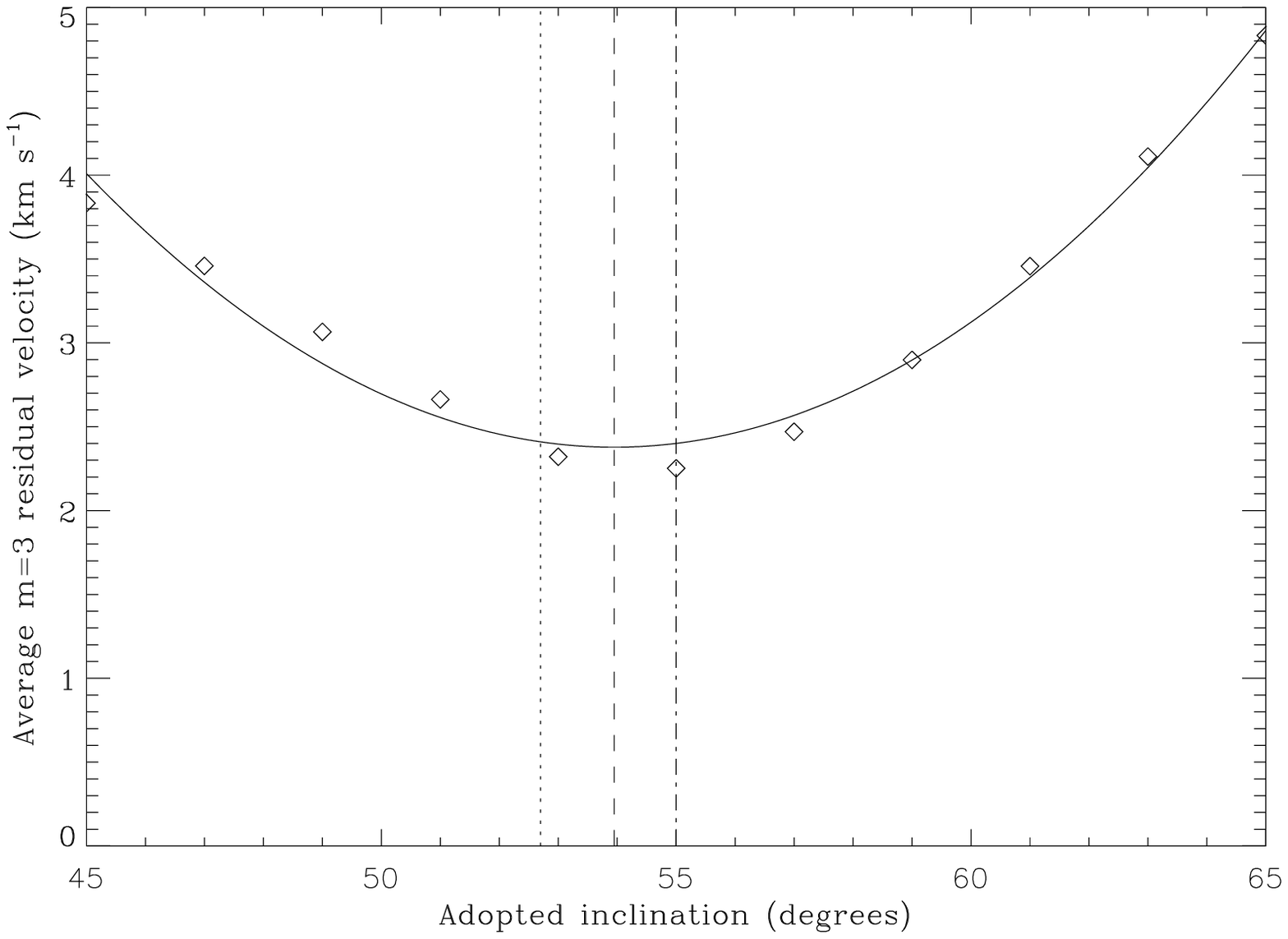]{Average amplitude of the $m=3$ component of the
deprojected velocity field of NGC~2915, as a function of the assumed
galactic inclination ($140\leq r\leq385\arcsec$). The solid line shows
a second order polynomial fit to the data, represented by
diamonds. The vertical lines mark the inclinations used in
Fig.~\ref{fig:m=3_rprofiles}.\label{fig:m=3_raverage}}
\figcaption[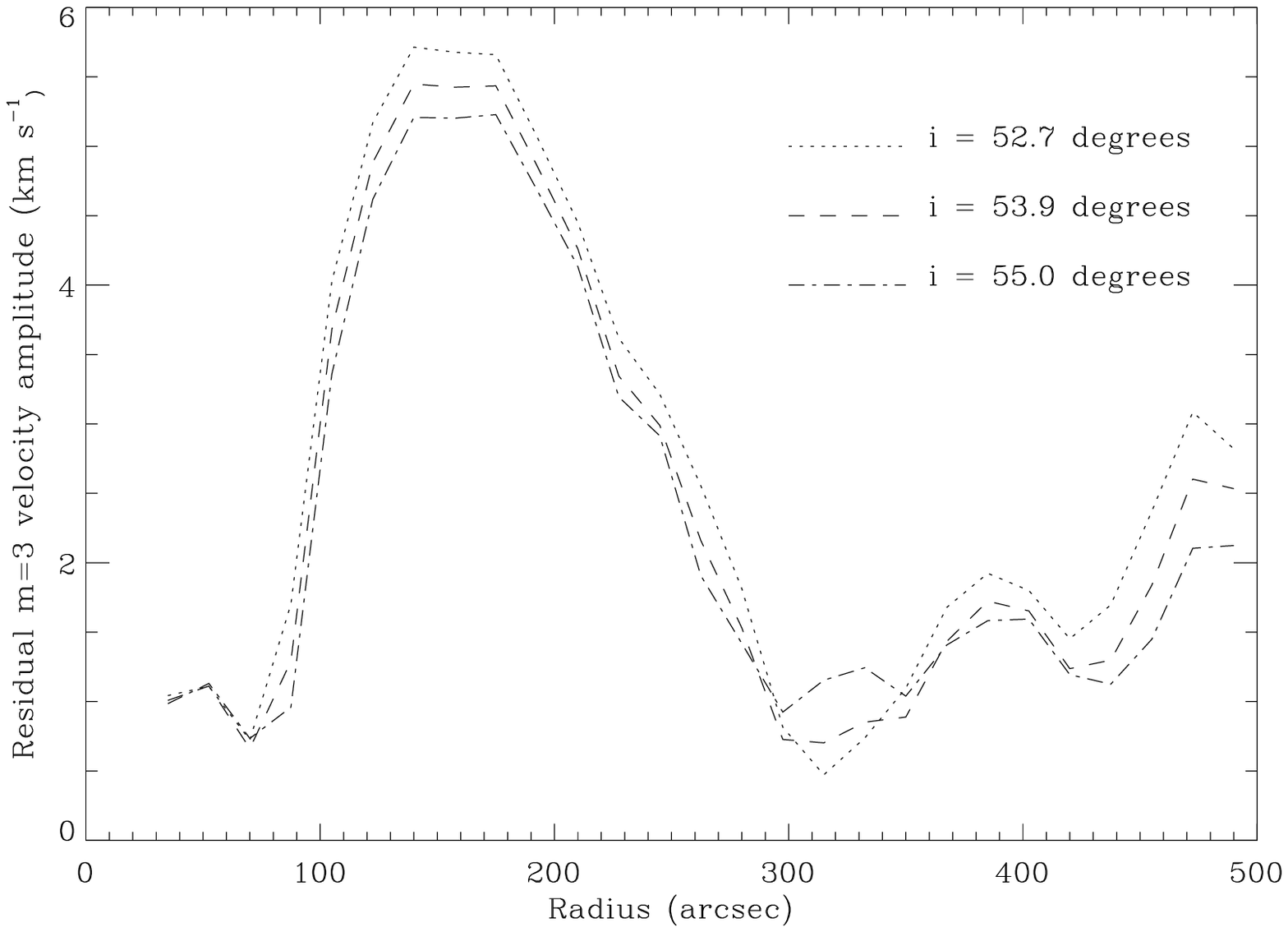]{Amplitude of the $m=3$ component of the
deprojected velocity field of NGC~2915, as a function of radius, for
three different values of the assumed galactic inclination. {\em From
top to bottom:} $i=$~55\fdg0, 53\fdg9, and 52\fdg7. The data are
truncated at half the beam size.\label{fig:m=3_rprofiles}}
\figcaption[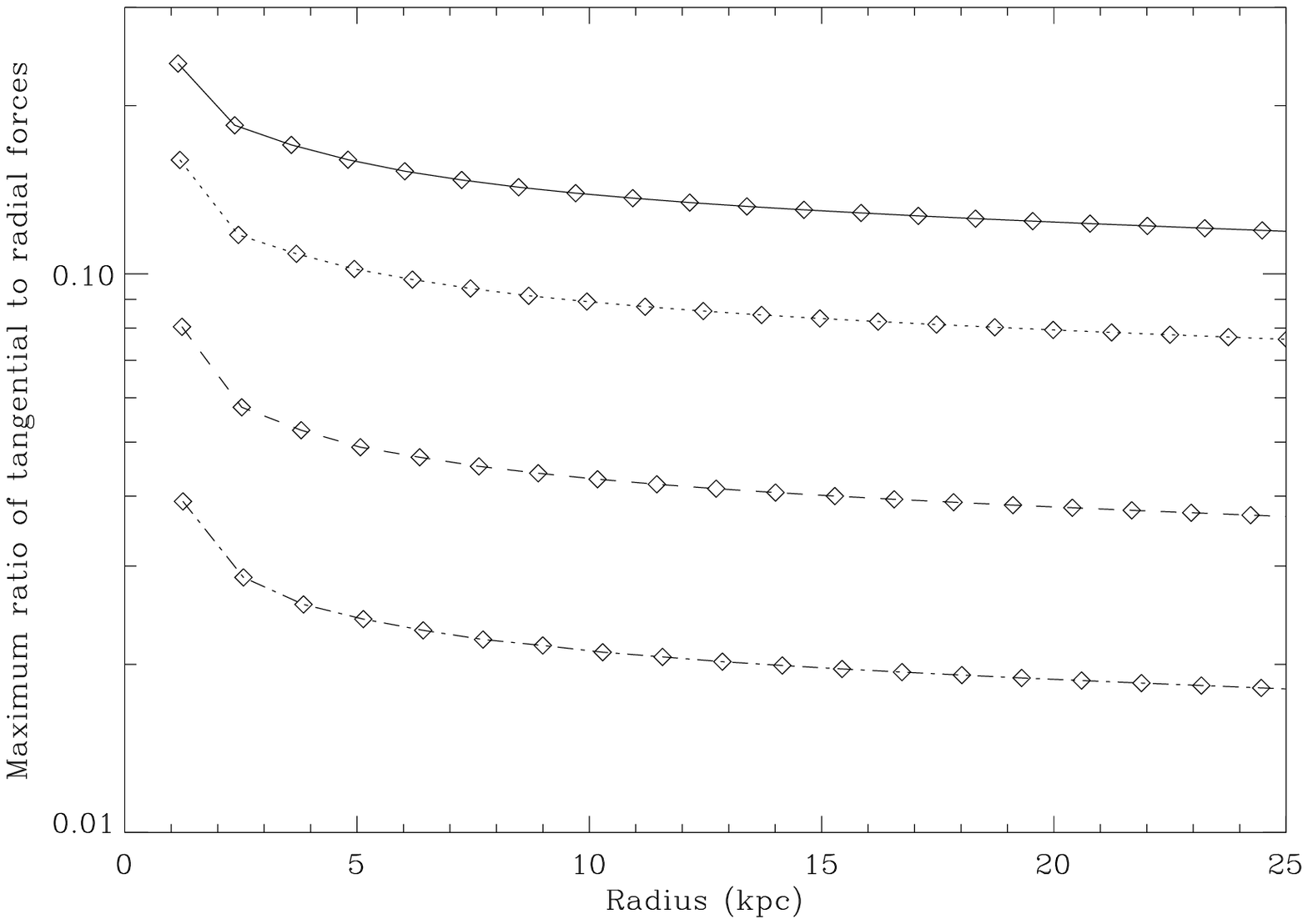]{Radial profile of the tangential to radial
acceleration ratio for four different triaxial halos. {\em From top to
bottom:} $b/a=0.7$, $0.8$, $0.9$, and $0.95$. All halos have $c/a=1$
and a core radius $r_c=2$~kpc.\label{fig:acc_ratio_profiles}}
\figcaption[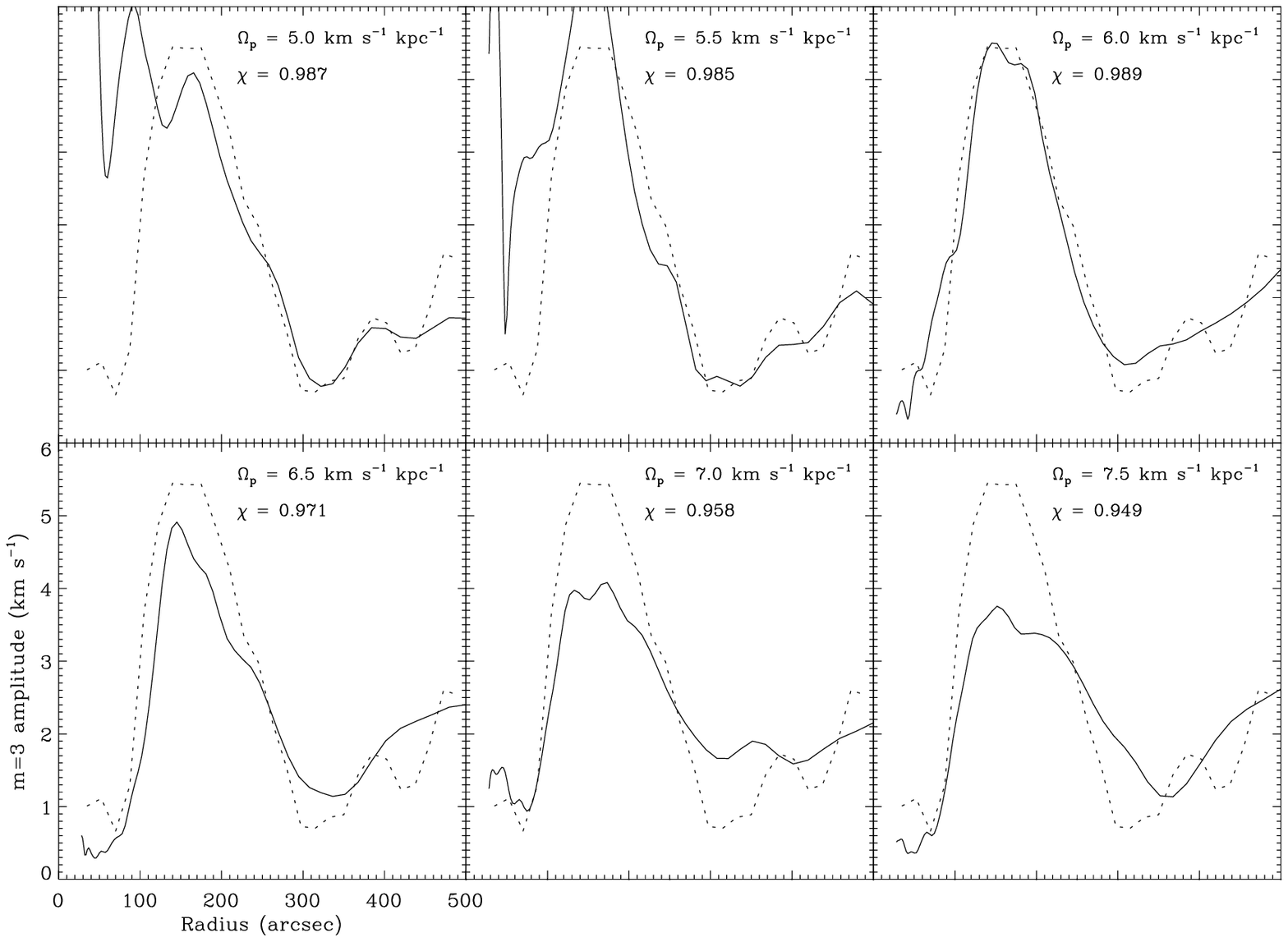]{Amplitude of the $m=3$ component of the
deprojected velocity field, as a function of radius for our best bar
forcing runs. Solid lines show the best match in each case; dotted
lines represent the observations. The pattern frequency $\Omega_p$ and
correlation coefficient $\chi$ are indicated in the top-right corner
of each panel. Contrary to Figs.~\ref{fig:barruns2}
and~\ref{fig:barruns3}, these data were not convolved by the
beam.\label{fig:barruns1}}
\figcaption[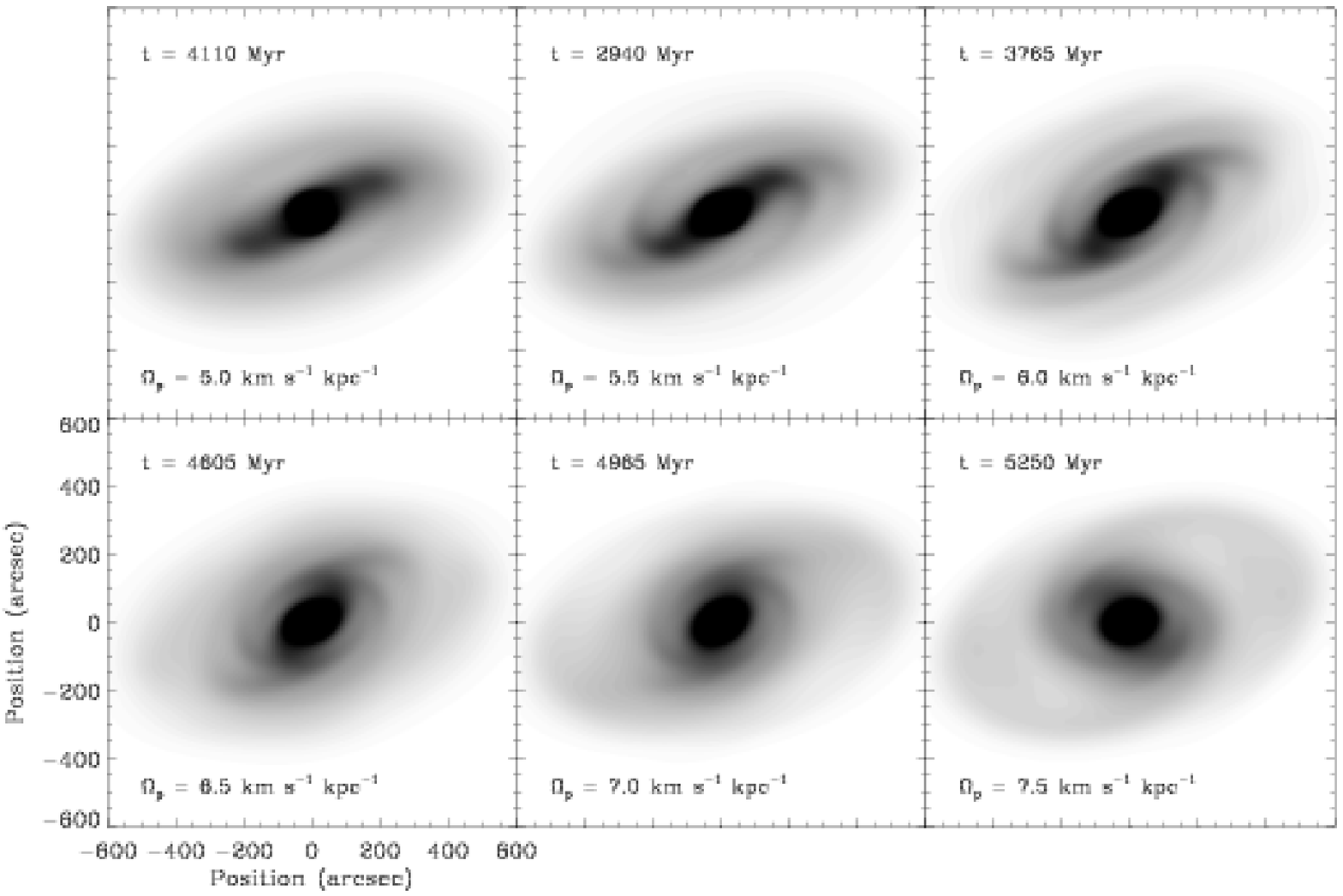]{Synthetic column density maps for our best bar
forcing runs (see Fig.~\ref{fig:barruns1}). The timestep and pattern
frequency are indicated in each panel. Convolution by a beam size of
$45\arcsec$ has been applied.\label{fig:barruns2}}
\figcaption[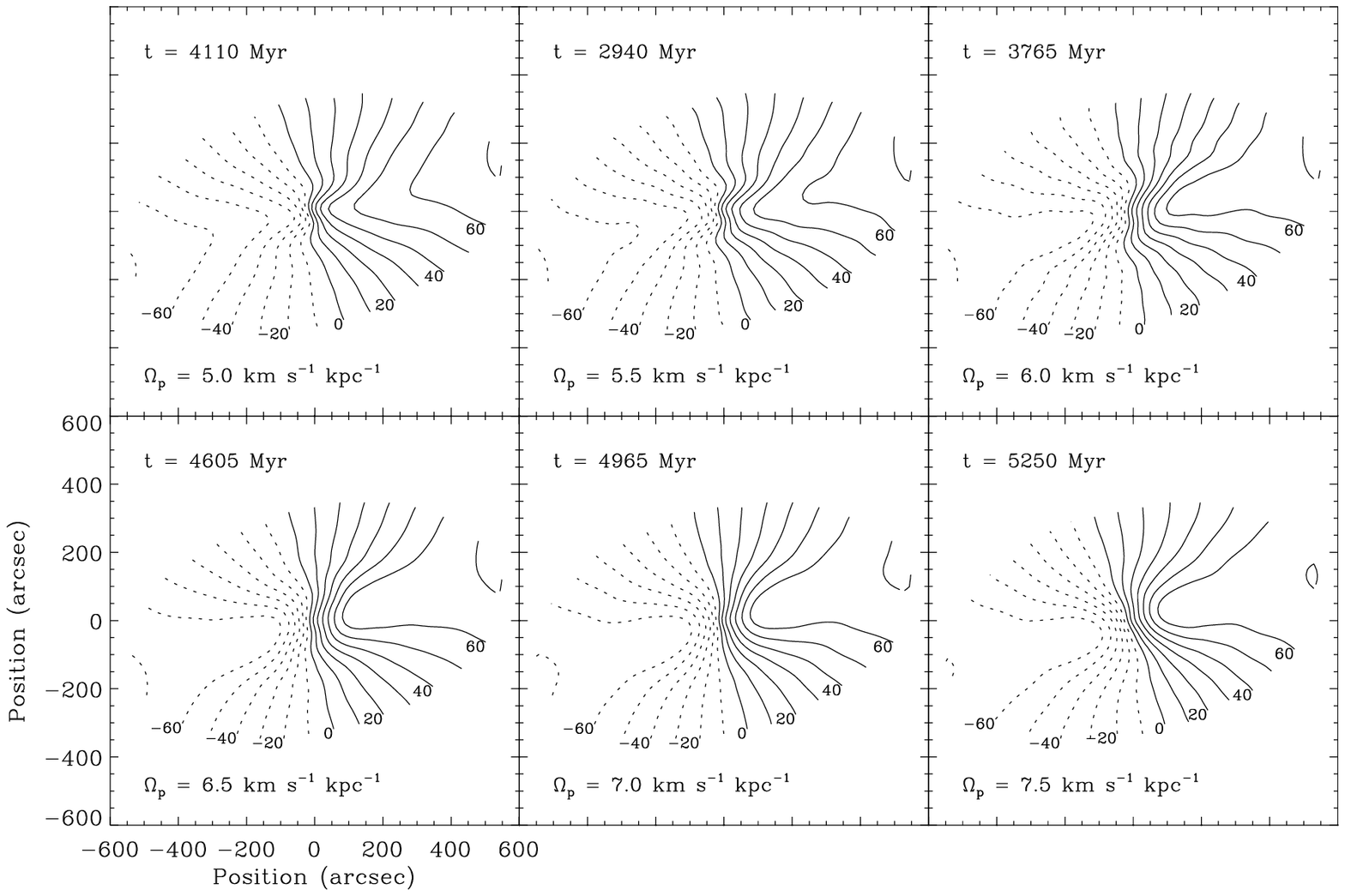]{Synthetic velocity fields for our best bar forcing
runs (see Figs.~\ref{fig:barruns1} and~\ref{fig:barruns2}). The
timestep and pattern frequency are indicated in each
panel. Convolution by a beam size of $45\arcsec$ has been
applied.\label{fig:barruns3}}
\figcaption[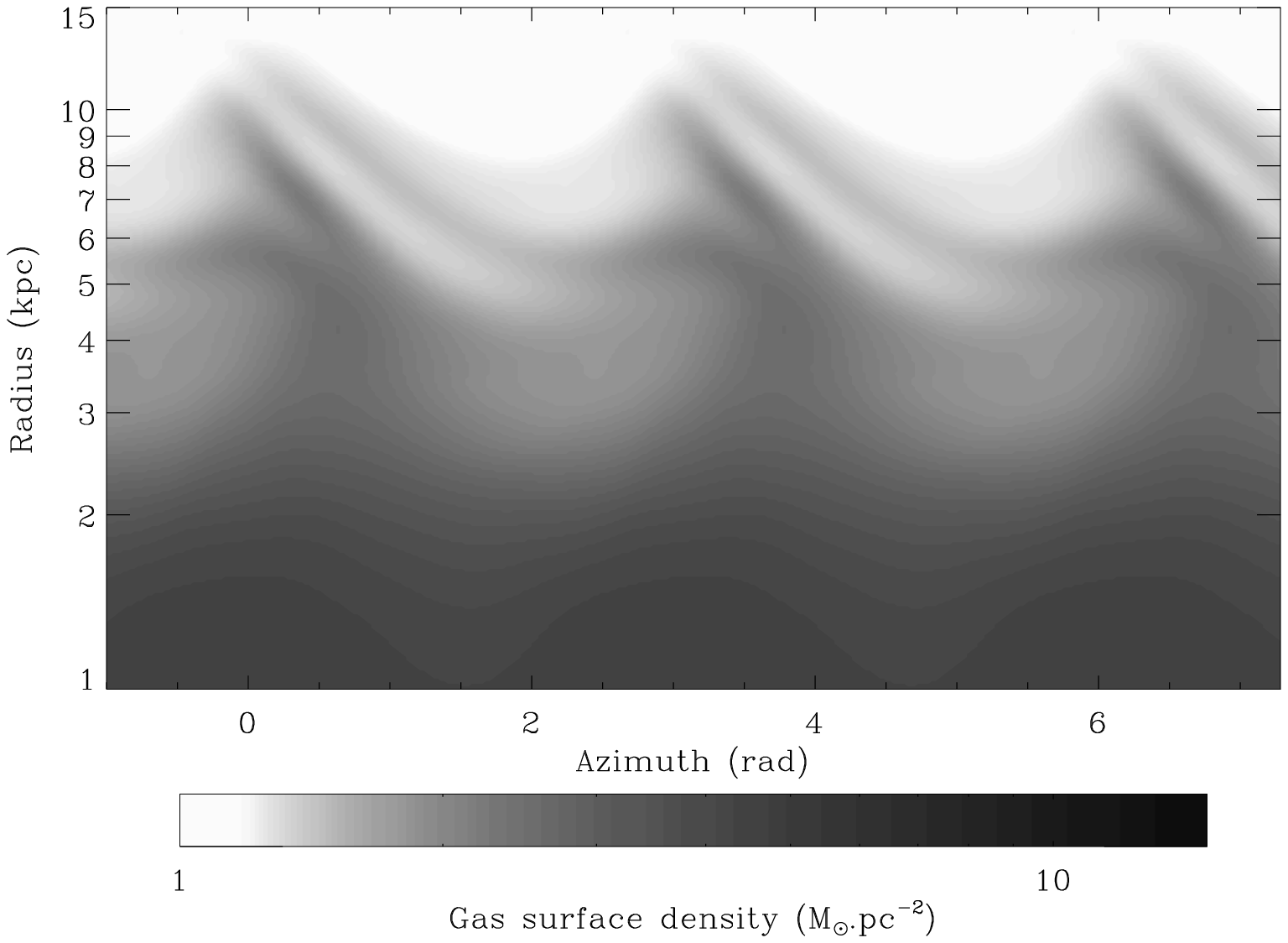]{Deprojected surface density of the bar forcing run
with $\Omega_p=6.0$~km~s$^{-1}$~kpc$^{-1}$, in $(\log r,\theta)$
coordinates, at the best match time. Convolution by a beam size of
$45\arcsec$ has been applied.\label{fig:logrtheta_barun}}
\figcaption[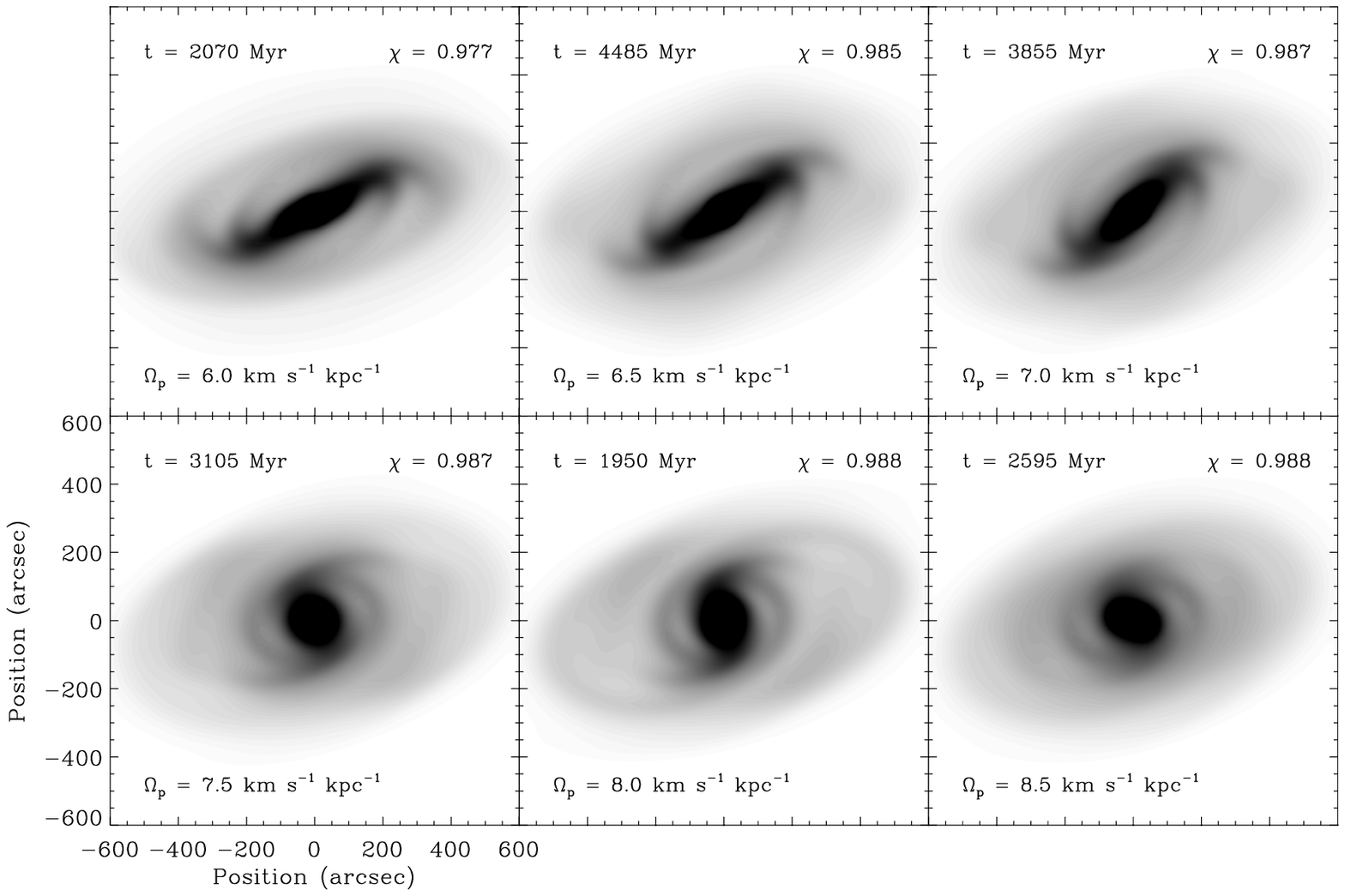]{Synthetic column density maps for our best halo
forcing runs The timestep and pattern frequency are indicated in each
panel. Convolution by a beam size of $45\arcsec$ has been
applied.\label{fig:halores1}}
\figcaption[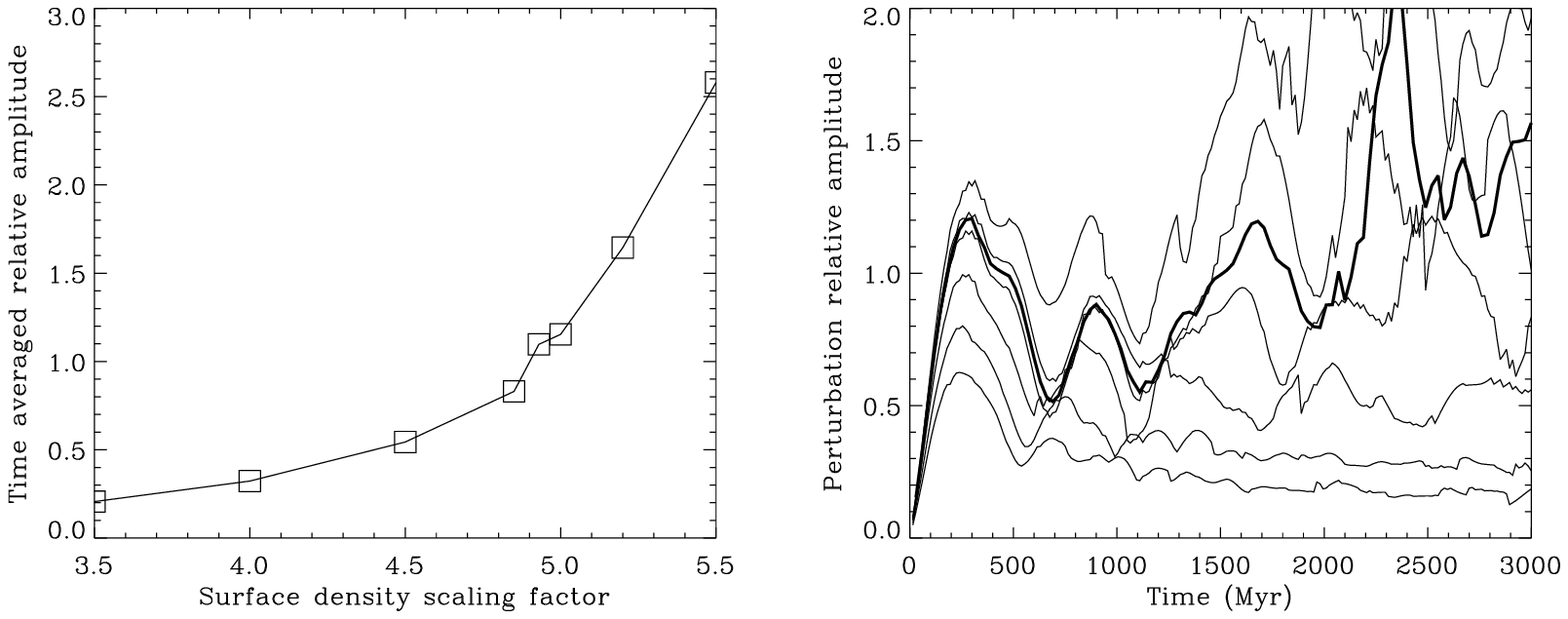]{Relative amplitude of the spiral instability for
the heavy disk models. {\em Left:} Time averaged relative amplitude
$\overline{\cal S}(\lambda_\Sigma)$ as a function of the surface
density scaling ratio $\lambda_\Sigma$. {\em Right:} Time dependence
of ${\cal S}(t)$ for various values of $\lambda_\Sigma$ (identified by
squares in the left panel). {\em From top to bottom:}
$\lambda_\Sigma=5.2$, $5.0$, $4.93$, $4.85$, $4.5$, $4.0$ and
$3.5$. The time behavior for $\lambda_\Sigma=5.5$ is not shown as the
disk is very close to $Q=1$ and fragments into two large symmetrical
clumps. The $\lambda_\Sigma=4.93$ curve, identified by a thick line,
is the best match model presented in this section and has the average
relative amplitude $\overline{\cal S}(\lambda_\Sigma)$ closest to
one. Despite the relatively large scatter in ${\cal S}(t)$, only a
narrow range of $\lambda_\Sigma$ can account for the observed
amplitude. Contrary to Figs.~\ref{fig:bestheavy}
and~\ref{fig:besthvlogrt}, these data were not convolved by the
beam.\label{fig:scaling}}
\figcaption[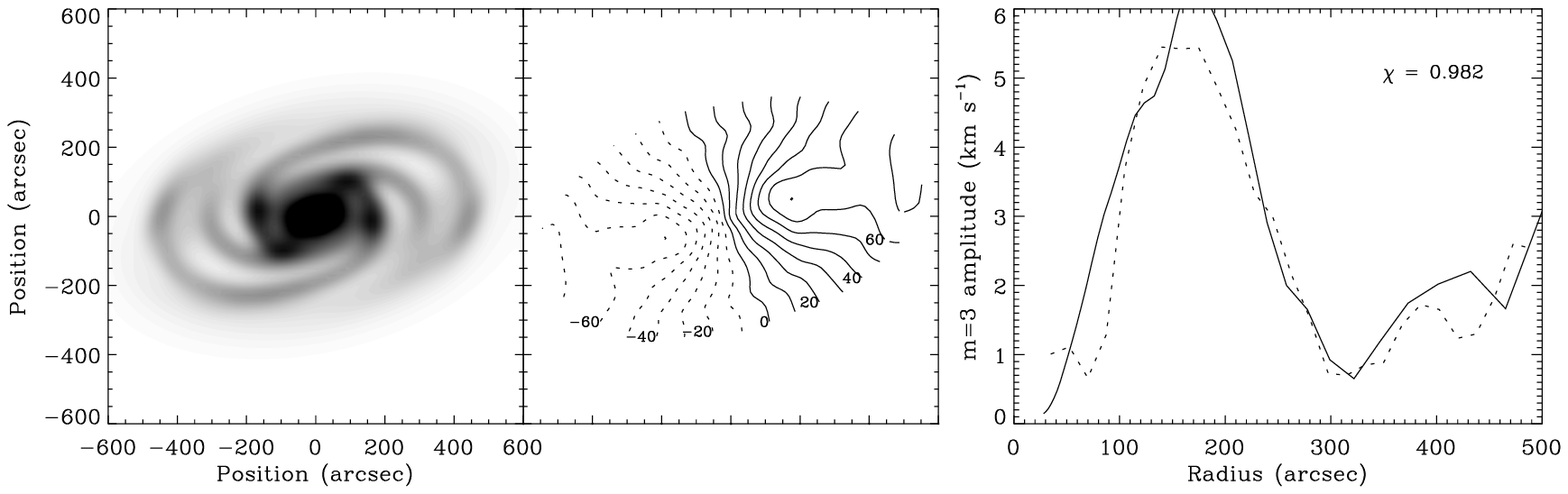]{Best match heavy disk model, with
$\lambda_\Sigma=4.93$ ($t=2.85$~Gyr). {\em Left:} Synthetic column
density map. {\em Center:} Synthetic velocity field. {\em Right:}
Amplitude of the $m=3$ component of the deprojected velocity field as
a function of radius. The solid line represents the model ($V_3^{\rm
sim}(r)$) and the dotted line the observations ($V_3^{\rm
obs}(r)$). Convolution by a beam size of $45\arcsec$ has been applied
to the first two plots only.\label{fig:bestheavy}}
\figcaption[fig15.eps]{Best match heavy disk model, with
$\lambda_\Sigma=4.93$ ($t=2.85$~Gyr), in $(\log r,\theta)$
coordinates. The applied correction factor $q=0.5$. Convolution by a
beam size of $45\arcsec$ has been applied.\label{fig:besthvlogrt}}
\figcaption[fig16.eps]{Halo and disk circular velocity and cumulative
mass profiles for our best match heavy disk model, with
$\lambda_\Sigma=4.93$ ($t=2.85$~Gyr). {\em Left:} Scaled gas disk
({\em dotted line}), dark halo ({\em dashed line}), and total ({\em
solid line}) circular velocity curves. {\em Right:} Disk ({\em solid
line}) and halo ({\em dotted line}) cumulative mass
profiles.\label{fig:massfraction}}
\figcaption[fig17.eps]{Surface density from our best match heavy disk
model with $\lambda_\Sigma=4.93$ ({\em solid line}), \citet{k89}
standard critical star formation threshold ({\em dashed line}), and
our modified critical star formation threshold ({\em dotted
line}). The observed velocity dispersion profile $\sigma_v(r)$ was
used to derive all the curves.\label{fig:keni}}
\clearpage
\begin{deluxetable}{cc}
\tablewidth{0pt}
\tablecaption{Bar masses for the runs presented in
Figs.~\ref{fig:barruns1}--\ref{fig:barruns3}. The other bar parameters
were kept fixed at $h_b=3$~kpc, $b/a=0.5$, and
$n=3$.\label{tab:barmasses}}
\tablehead{\colhead{\coltwo{$\Omega_p$}{(km~s$^{-1}$~kpc$^{-1}$)}} & 
           \colhead{\coltwo{$M_b$}{($10^9$~\msun)}}}
\startdata
5.0 & 4.9\\
5.5 & 5.7\\
6.0 & 6.8\\
6.5 & 5.1\\
7.0 & 4.8\\
7.5 & 4.7\\
\enddata
\end{deluxetable}
\clearpage
\begin{figure}
\epsscale{0.6}
\plotone{f1.eps}
\end{figure}
\clearpage
\begin{figure}
\epsscale{1.0}
\plotone{f2.eps}
\end{figure}
\clearpage
\begin{figure}
\plotone{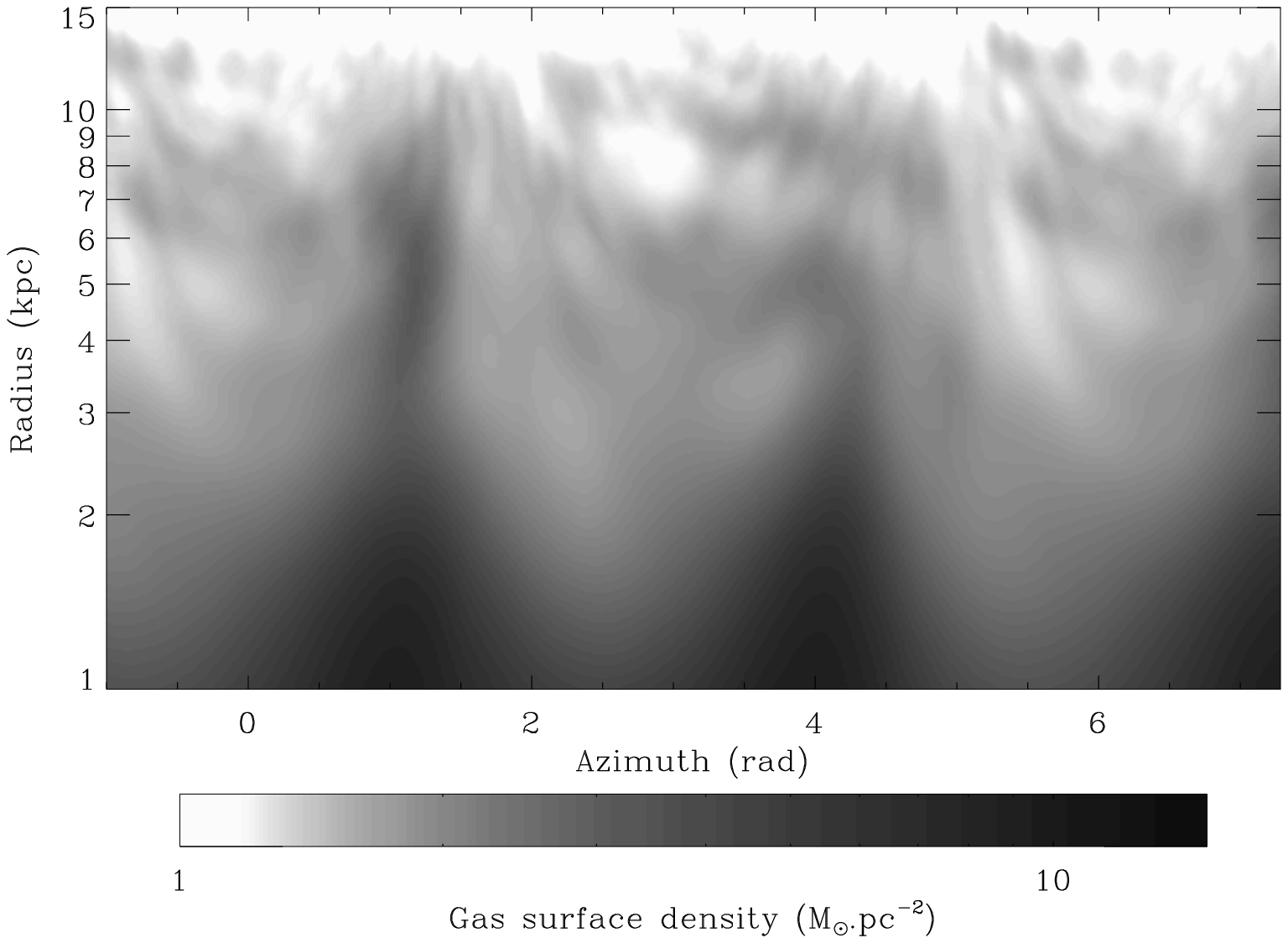}
\end{figure}
\clearpage
\begin{figure}
\plotone{f4.eps}
\end{figure}
\clearpage
\begin{figure}
\plotone{f5.eps}
\end{figure}
\clearpage
\begin{figure}
\plotone{f6.eps}
\end{figure}
\clearpage
\begin{figure}
\plotone{f7.eps}
\end{figure}
\clearpage
\begin{figure}
\plotone{f8.eps}
\end{figure}
\clearpage
\begin{figure}
\plotone{f9.eps}
\end{figure}
\clearpage
\begin{figure}
\plotone{f10.eps}
\end{figure}
\clearpage
\begin{figure}
\plotone{f11.eps}
\end{figure}
\clearpage
\begin{figure}
\plotone{f12.eps}
\end{figure}
\clearpage
\begin{figure}
\plotone{f13.eps}
\end{figure}
\clearpage
\begin{figure}
\plotone{f14.eps}
\end{figure}
\clearpage
\begin{figure}
\plotone{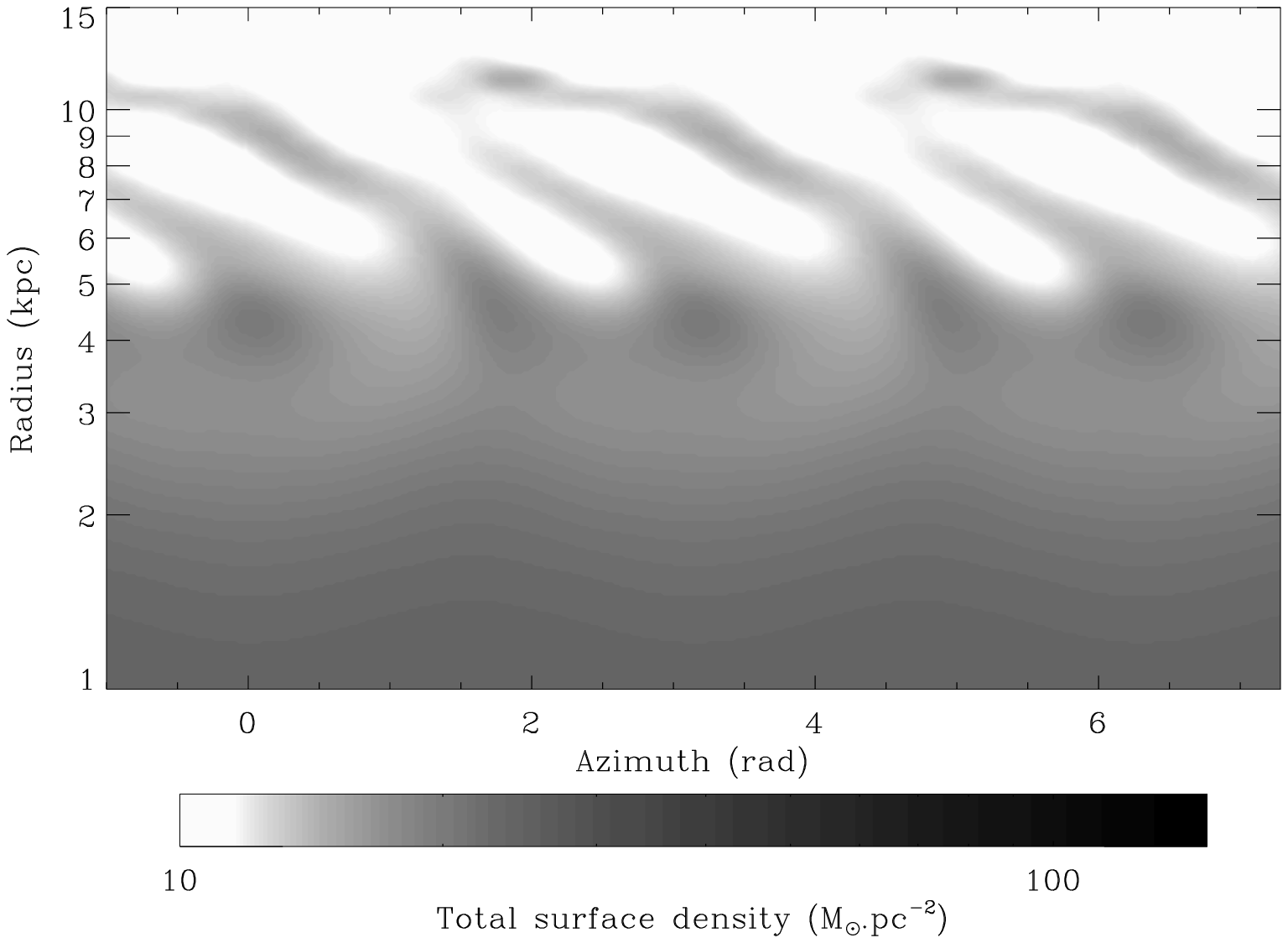}
\end{figure}
\clearpage
\begin{figure}
\plotone{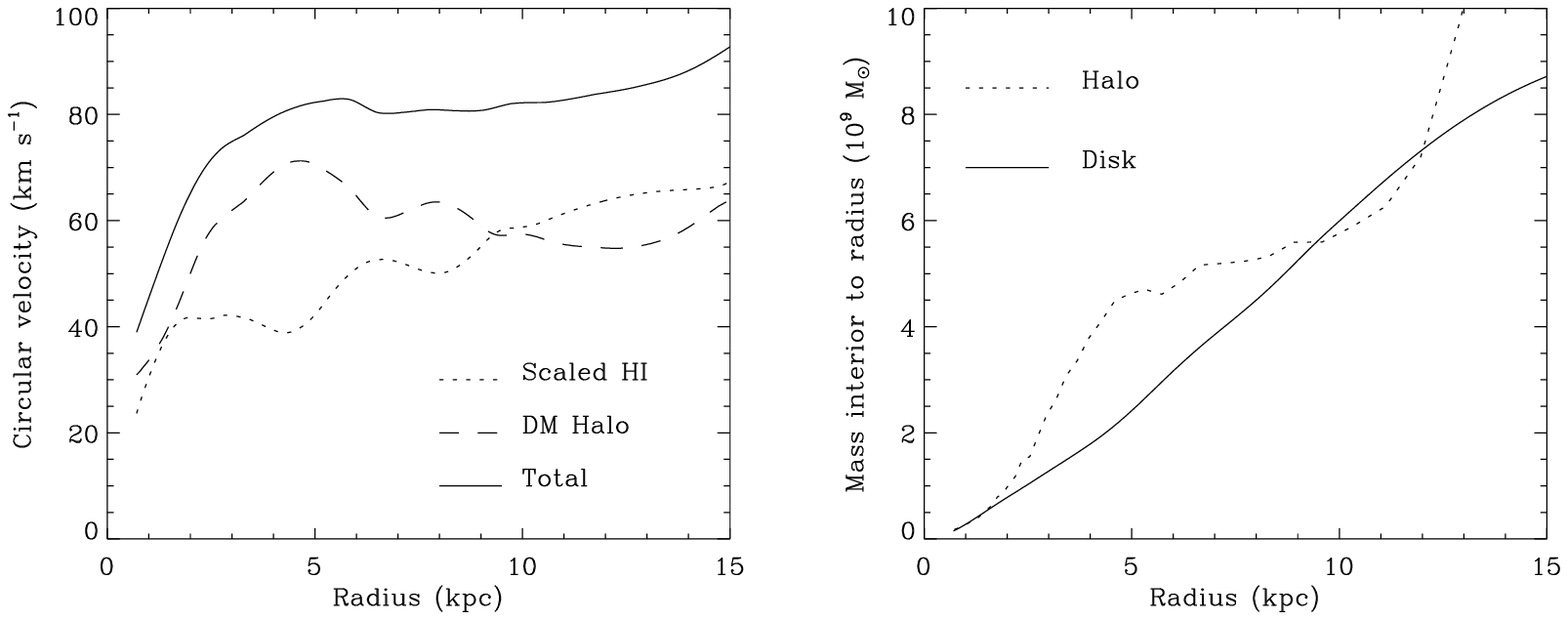}
\end{figure}
\clearpage
\begin{figure}
\plotone{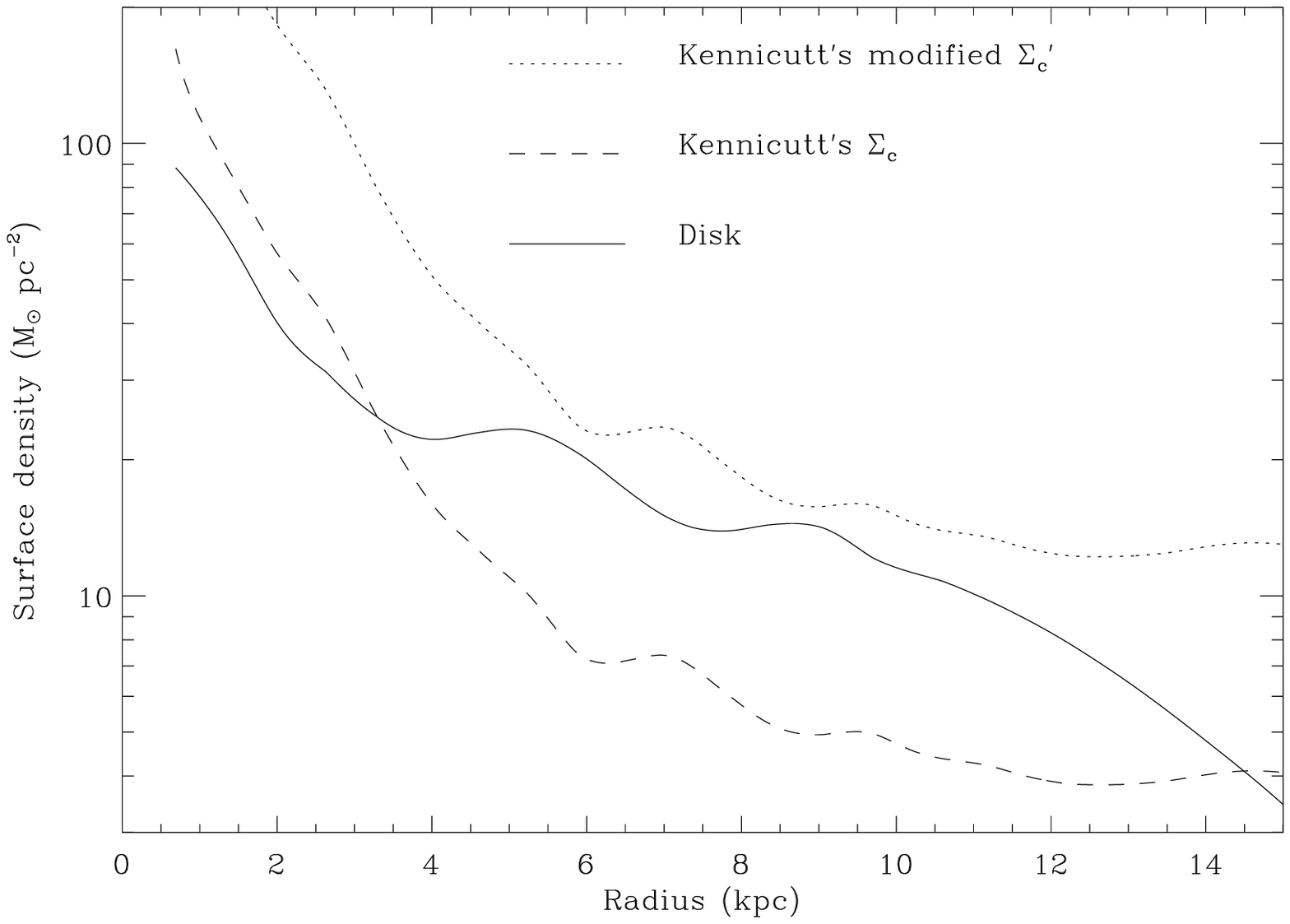}
\end{figure}
\end{document}